\documentclass{aastex}
\usepackage{epsf}
\usepackage{emulateapj5}
\usepackage{onecolfloat}

\newcounter{affil}

\def\lesssim{\lower0.6ex\vbox{\hbox{$ \buildrel{\textstyle <}\over{\sim}\ $}}}
\def\gsim{\lower0.6ex\vbox{\hbox{$ \buildrel{\textstyle >}\over{\sim}\ $}}}

\def\gtorder{\mathrel{\raise.3ex\hbox{$>$}\mkern-14mu
             \lower0.6ex\hbox{$\sim$}}}
\def\ltorder{\mathrel{\raise.3ex\hbox{$<$}\mkern-14mu
             \lower0.6ex\hbox{$\sim$}}}

\def\beq{\begin{equation}}
\def\eeq{\end{equation}}

\def\beqa{\begin{eqnarray}}
\def\eeqa{\end{eqnarray}}

\def\bfig{\begin{figure}[t] \begin{center}}
\def\efig{\end{center} \end{figure}}

\def\bfignoc{\begin{figure}[t]}
\def\efignoc{\end{figure}}

\begin{document}

%\slugcomment{{\em Astrophysical Journal, submitted}}

%\onecolumn  (Can be used WITHOUT [] to create a single column environment elsewhere in the document)

\twocolumn[%%% Begin one-column header material

\slugcomment{{\em }}

\lefthead{X-ray Evolution of AGES Galaxies}
\righthead{Watson et al.}

\title{The Star Formation and Nuclear Accretion Histories of Normal Galaxies in the AGES Survey}
\author{
Casey R. \ Watson\altaffilmark{1},
Christopher S.\ Kochanek\altaffilmark{2}, 
William R. \ Forman\altaffilmark{3}, 
Ryan C. Hickox\altaffilmark{3},
Christine J. \ Jones\altaffilmark{3},
Michael J. I. \ Brown\altaffilmark{4},
Kate \ Brand\altaffilmark{5},
Arjun \ Dey\altaffilmark{6}, 
Buell T. \ Jannuzi\altaffilmark{6},
Almus T. \ Kenter\altaffilmark{3}, 
Steve S. \ Murray\altaffilmark{3},
Alexey \ Vikhlinin\altaffilmark{3}, 
Daniel J. \ Eisenstein\altaffilmark{7}, 
Giovani G. \ Fazio\altaffilmark{3},
Paul J. \ Green\altaffilmark{3}, 
Brian R. \ McNamara\altaffilmark{8},  
Marcia \ Rieke\altaffilmark{6}, 
Joseph C.\ Shields\altaffilmark{8}}

\begin{abstract}

We combine IR, optical and X-ray data
from the overlapping, 9.3 square degree NOAO Deep Wide-Field Survey (NDWFS),
AGN and Galaxy Evolution Survey (AGES), and XBo\"otes Survey to measure
the X-ray evolution of 6146 normal galaxies as a function of 
absolute optical luminosity, redshift, and
spectral type over the largely unexplored redshift range $0.1 \ltorder z \ltorder 0.5$. 
Because only the closest or brightest of the galaxies are individually detected in X-rays,
we use a stacking analysis to determine the mean properties of the sample.  
Our results suggest that X-ray emission from spectroscopically 
late-type galaxies is dominated by star formation, while that from
early-type galaxies is dominated by a combination of hot gas and AGN emission.  
We find that the mean star formation and supermassive black 
hole accretion rate densities evolve like $\sim (1+z)^{3\pm 1}$, in agreement
with the trends found for samples of bright, individually detectable starburst
galaxies and AGN. Our work also corroborates
the results of many previous stacking analyses of
faint source populations, with improved statistics.

\end{abstract}

\keywords{
  cosmology: observations
  ---
  galaxy evolution
 }

]%%% end twocolumn header

\altaffiltext{1}{Department of Physics and Astronomy, Millikin University, 1184 West Main Street,
Decatur, IL 62522;\\ crwatson@millikin.edu}

\altaffiltext{2}{Department of Astronomy, The Ohio State University, 140 West 18th
Avenue, Columbus, OH 43210}

\altaffiltext{3}{Harvard-Smithsonian Center for Astrophysics, 60 Garden Street,
Cambridge, MA 02138}

\altaffiltext{4}{Princeton University Observatory, Peyton Hall,
                Princeton, NJ 08544}

\altaffiltext{5}{Space Telescope Science Institute, 3700 San Martin Drive, Baltimore, MD 21218}

\altaffiltext{6}{National Optical Astronomy Observatory, Tucson, AZ 85726-6732}

\altaffiltext{7}{Steward Observatory, University of Arizona, 933 North Cherry
Avenue, Tucson, AZ 85721}

\altaffiltext{8}{Department of Physics and Astronomy, Ohio University, Athens, OH
45701}

\section{INTRODUCTION}

There are three primary sources of galactic X-ray emission: diffuse, hot gas, 
accreting stellar remnants, such as X-ray binaries,
and accreting supermassive black holes, i.e., active galactic nuclei (AGN).
Apart from nearby ellipticals, which tend to be dominated by hot gas emission
(e.g., Forman, Jones, \& Tucker 1994), local surveys indicate that
X-ray binaries dominate the flux from ``normal'' galaxies with 
quiescent nuclei (e.g., 
Fabbiano \& White 2003, Smith \& Wilson 2003, Muno et al. 2004). 
There are two distinct populations of X-ray binaries: the short-lived 
($\ltorder 10^6$ yrs), high-mass X-ray binaries (HMXBs) and the long-lived
($\gtorder 10^8$ yrs), low-mass X-ray binaries (LMXBs). 
Because of their short lifetimes, one expects
the X-ray emission from HMXBs to track the current star formation
rate (SFR).  In contrast, the 
X-ray emission from longer-lived LMXBs should track the
integrated stellar mass with some time lag due to stellar and binary
evolution (e.g., Ghosh \& White 2001, Ptak et al. 2001, Grimm et al. 2002).

These expectations are largely borne out by observations.  There are
strong correlations between the hard ($> 2$ keV) X-ray luminosity (presumably 
from HMXBs) and other star-formation
indicators, such as radio, far-IR, B-band, and UV flux 
for both local (e.g.,
Fabbiano 1989, David, Jones, \& Forman 1992) and higher redshift (Seibert et al. 2002, Bauer et al. 
2002) galaxy samples.
Many studies have found a linear correlation between X-ray luminosity and the SFR
(e.g., Ranalli et al. 2002, Grimm et al. 2003, Gilfanov et al. 2004, Persic et al. 2004, 
and Colbert et al. 2004).  In particular, Grimm et al. (2003) found that the hard 
(2--10~keV) HMXB X-ray luminosity scales with the 
star formation rate according to
\begin{equation}
        L^{\rm HMXB}_{\rm x,hard} \simeq 0.67\times 10^{40}\left( { \hbox{SFR} 
\over M_\odot \hbox{yr}^{-1} } \right)~\rm{ergs~s}^{-1}.
\label{eq:LxSFR}
\end{equation}
In contrast, the (total band: 0.5--8 keV) X-ray emission from LMXBs
is well-correlated with the $K$-band flux of galaxies, which in turn is a good tracer of 
stellar mass.  
Kim and Fabbiano (2004) find that 
\begin{equation}
        L^{\rm LMXB}_{\rm x} \simeq 10^{40}\left( { L_K \over L_{K*} } \right)~\rm{ergs~s}^{-1},  
\label{eq:LxLK}
\end{equation} 
where $L_{K*} = \nu_K L_{\nu_K , *} = 2.6\times10^{43}$~ergs~s$^{-1}$ corresponds to $M_{K*}=-23.4$~mag (Kochanek et al. 2001). 

Because the stellar mass
density evolves slowly between $z=0$ and $z \simeq 0.5$ 
(e.g., Bell 2004 and references therein), we 
expect little change in the number of LMXBs and assume that Eqn.~(\ref{eq:LxLK}) holds out to $z \simeq 0.5$.
On the other hand, the star formation rate density rises rapidly with look back time, (e.g.,
$\propto (1+z)^{2.7 \pm 0.7}$ Hogg 2001), so
we should see a
correspondingly rapid evolution in the X-ray flux of star-forming galaxies due to 
the increasing number of HMXBs.  

The accretion luminosity density of AGN also rises rapidly with redshift 
(e.g., $\propto (1+z)^{3.2 \pm 0.8}$, Barger et al. 2005) with
higher densities of more luminous sources at higher redshifts
(e.g., Barger et al. 2001, Cowie et al. 2003, Ueda et al. 2003,
Miyaji 2004, Hasinger 2004, Hasinger, Miyaji, \& Schmidt 2005).  
The observed trends suggest that faint AGN with characteristic X-ray luminosities 
approaching those of bright starbursts, $L_{\rm x} \ltorder 10^{41} ~\hbox{ergs s}^{-1}$, 
may be most abundant at $z\ltorder 1$ and, if so, they could produce
a substantial fraction of the ``normal'' galaxy X-ray flux at these 
redshifts.  

Given their similar evolution, spectral shapes
(e.g., Ptak et al. 1999), and 
potentially similar luminosities,
it may be difficult to disentangle the X-ray emission from
AGN and star formation at $z \ltorder 1$.
Probing the X-ray evolution of normal galaxies between $z \simeq 0.1$ and $z \simeq 1$ 
is also difficult because it
requires relatively deep observations over relatively wide areas.
In fact, most of our knowledge about the X-ray 
properties of galaxies still comes from either large-area,
local surveys or high-redshift, small-volume deep fields. 

This is the second in a series of papers in which we attempt to bridge this gap by examining 
the X-ray evolution of galaxies within the $9.3$~square degree Bo\"otes field of the NOAO
Deep Wide-Field Survey (NDWFS; Jannuzi \& Dey 1999, Jannuzi et al., in preparation, 
Dey et al. in preparation). 
The XBo\"otes survey (Murray et al. 2005)
obtained a 5 ksec Chandra mosaic of the entire $9.3$~square degree field based on 126 ACIS-I pointings.
These data are too shallow, however, to detect typical galaxies even at modest redshifts. 
Fortunately, the wide area of the survey allows us to measure the mean properties
of a large, representative population of galaxies by ``stacking'' (averaging) their X-ray emission 
(Brandt et al. 2001; Nandra et al. 2002; Hornschemeier et al. 2002,
Georgakakis et al. 2003, Lehmer et al. 2005, Laird et al. 2005, Laird et al. 2006, Lehmer et al. 2007).   

In the first paper in the series, 
Brand et al.~(2005; hereafter B05) employed the stacking technique to examine the
X-ray luminosity evolution of $\approx 3300$ optically luminous ($\sim L_*$),
red galaxies with photometric
redshifts $0.3 < z < 0.9$.  By constraining the sample to have the same
evolution-corrected, absolute $R$-band magnitude distribution at all redshifts
($M_R < -21.3$, $\langle{M}_{R}\rangle \simeq -22.0$), 
it was possible to follow the nuclear accretion histories of a consistent 
population of galaxies throughout this epoch. 
Because these massive, early-type
galaxies are the typical hosts of powerful quasars at higher redshifts
(McLeod \& McLeod 2001; Dunlop et al. 2003), 
tracking the decay of their nuclear activity to lower redshifts
is essential to our understanding of the decline in supermassive black hole (SMBH) accretion from its peak during the
quasar phase ($z \gtorder 2$) to the present.
The observed variation in nuclear accretion rate of the $>L_*$ red galaxies,
$L_{\rm x} \propto (1+z)^{4\pm2.4}$,
is broadly consistent with the behavior of bright, individually detected AGN
Barger et al. (2001; hereafter Ba01) and Barger et al. (2005; hereafter Ba05).
 
In the present paper we combine the NDWFS and XBo\"otes surveys with
redshifts from the AGN and Galaxy Evolution Survey (AGES, Kochanek et al. 2007) 
to study the X-ray properties of a complete, flux-limited
sample of galaxies as a function of absolute optical luminosity, redshift, stellar mass, and spectral
type.  While our general procedures are similar to those of other recent
stacking analyses (e.g., 
Brandt et al. 2001; Nandra et al. 2002; Hornschemeier et al. 2002; 
Georgakakis et al. 2003, Laird et al. 2005, Laird et al. 2006, Lehmer et al. 2007), the number 
of objects we consider ($\approx 6500$) is 10--100 times larger.
The size of our sample allows us to convert the short mean exposure time of 
the XBo\"otes survey (5~ksec) into stacked, effective exposure times that are large
enough to detect galaxies at intermediate redshift (Msec).  

In \S2 we describe the data and illustrate our ability to 
detect X-ray emission from the AGES galaxies through the stacking approach.  
In \S3 we provide further details of our stacking analysis and use Monte Carlo simulations
to test it and to refine our choices of signal and background apertures. In \S4
we present measurements of the mean evolution of all the AGES galaxies. We then consider
the radial emission profiles of a nearby ($z \leq 0.1$) subsample in order to test for 
weak/obscured AGN. 
We also examine the hardness and  
X-ray to optical luminosity ratios as a function of redshift, 
stellar mass, and
spectroscopic type (late-type, early-type, or AGN). 
Based on these tests, we evaluate the relative contributions of
hot gas, LMXBs, HMXBs, and AGN emission to the observed signal. 
We then discuss our results and compare them to previous studies.
Lastly, we determine the evolution of the star 
formation rate per unit galactic stellar mass and the star formation rate density based on 
emission from the late-type galaxy sample and trace 
the evolution of the nuclear accretion rate and accretion rate density of 
AGN in the early-type galaxy sample.  
We summarize our findings in \S5.  
Throughout the paper, we assume a flat $\Lambda$CDM cosmology 
with $\Omega_{\rm m,0}=0.3$, $\Omega_{\Lambda ,0}=0.7$ and $H_0=70~ \hbox{km s}^{-1} ~\rm  Mpc^{-1}$.

\section{The X-ray Data, The Galaxy Sample and Methods}

The NDWFS obtained $B_W$, $I$, $R$, and
(partial) $K$-band observations of two roughly $9.3$ square degree regions.  
The Northern Bo\"otes field has also been surveyed at 
radio (VLA FIRST, Becker et al. 1995 and WSRT, de Vries et al. 2002),
far-IR (Spitzer/MIPS, Soifer et al., 2004),
mid-IR (Spitzer/IRAC, The IRAC Shallow Survey, Eisenhardt et al., 2004), 
near-IR (NDWFS and the Flamingos Extragalactic Survey; Elston et al. 2006), 
UV (GALEX; Hoopes 2004), and 
X-ray (XBo\"otes, Murray et al. 2005) wavelengths.
In particular, the XBo\"otes survey covered the entire Northern Bo\"otes field
with 126 $17'\times 17'$ ACIS-I images of mean exposure time 5~ksec using the Chandra X-ray Observatory (Murray et al. 2005).
The survey detected approximately 3300 point sources and 43 extended sources in the field with $\geq 4$ counts (4767 
sources with $\geq 2$ counts, Kenter et al.~2005).  

\subsection{X-ray Data}

For our current analysis we use the positions of the 420,000 photons detected 
in the XBo\"otes observations, with weights correcting
for off-axis vignetting and variations in exposure time, 
their estimated energies $\epsilon \pm \delta\epsilon$, and their positions
relative to the pointing center of the observations.  We limited the analysis
to photons with $0.5 < \epsilon < 7$~keV, dividing them
into total ($0.5-7$~keV), soft ($0.5-2$~keV) and hard ($2-7$~keV) energy bands. 
The mean total, hard, and soft backgrounds 
measured during these observations were approximately $2.8$, $1.9$ and 
$0.96\times 10^{-3}$ counts arcsec$^{-2}$, respectively.  

For the normal galaxies, the average ratio between 
the hard and soft X-ray counts is $0.73$.  To fit this
count ratio, we assume an absorbed power-law $dN(E)/dE \propto
E^{-\Gamma}\rm{~(counts~s^{-1}~keV^{-1})}$
for the intrinsic X-ray spectrum. We then use 
Portable, Interactive Multi-Mission Simulator
(PIMMS v3.9d)\footnote{http://heasarc.nasa.gov/Tools/w3pimms.html} to find the best-fit
photon index, $\Gamma = 1.29$, assuming a fairly low absorption
column density ($N(H) = 4 \times 10^{20}$~cm$^{-2}$), half of
which is the Galactic absorption for the B\"ootes field from
Stark et al.~(1992). Factor of two changes in the 
assumed column densities change $\Gamma$ by only a few percent. 
The spectrum of the ``normal'' galaxies is very similar to the 
that of the Cosmic X-ray Background, $\Gamma_{\rm CXB} \simeq 1.4$ (e.g., Tozzi et al. 2001;
Nandra et al. 2004)
and of resolved populations in the Chandra Deep Fields (e.g., Civano, Comastri, \& Brusa 2005). 
Photon indices in this range are
also typical of both X-ray binaries and obscured AGN 
($1 \ltorder \Gamma \ltorder 2$, e.g., Ptak et al. 1999, Muno et al. 2004).
Based on our model of the composite AGES galaxy X-ray spectrum, we estimate that one soft,
hard or total band count corresponds to  
$1.05 \times 10^{-15}$~ergs~cm$^{-2}$~s$^{-1}$, $3.65 \times 10^{-15}$~ergs~cm$^{-2}$~s$^{-1}$ and
$2.14 \times 10^{-15}$~ergs~cm$^{-2}$~s$^{-1}$, respectively, for
an on-axis source with a 5~ksec exposure time.

\subsection{The Galaxy Sample}

For our stacking analysis we cross-referenced the XBo\"otes 
catalog of X-ray photon positions (i.e., the event list) with those of the galaxies
observed by the AGN and Galaxy Evolution Survey (AGES, Kochanek et al. 2007).
We focus on the approximately 6500 galaxies in the 2004 AGES main galaxy sample.
This sample consists of all galaxies with $R<19.2$~ mag and a randomly 
selected 20\% of galaxies with $19.2<\hbox{R}<20$~mag.  We use Vega
magnitudes throughout the paper.
The spectroscopic redshifts of the galaxies were measured using Hectospec, a
robotic 300 fiber spectrograph for the MMT (Fabricant et al. 1998, Roll et al. 1998, 
Fabricant et al. 2005) and are 95\%
complete for $R<19.2$;
our analysis is not significantly affected by the modest level of incompleteness. 
Although we have redshifts
for roughly half of the remaining 80\% of galaxies with $19.2 < R < 20$~mag,
we do not include them in this analysis because they were
not targeted based on a simple R-band flux-limited criterion.

Redshifts were measured using two
independent data reduction pipelines (the Hectospec pipeline
at the CfA and a modified SDSS pipeline at Steward Observatory),
and the results were verified by visually inspecting the
spectra for features consistent with the pipeline redshift.
Both the pipelines and the visual inspection flagged galaxies
with spectroscopic signatures of AGN, both obvious broad-lined
systems and those with emission line ratios
characteristic of Type II optical AGN. We also carried out
a principal component analysis of the optical spectra
(e.g., Formiggini \& Brosch 2004, Ferreras et al. 2006).
In our analysis, the level of star formation was largely characterized by the
ratio of the second component relative to the first,
so we divided the sample into early and late-type galaxies
at the median of the distribution of that ratio for the
full AGES galaxy sample.

Rest-frame absolute $R$-band magnitudes, $M_R$, were determined by combining the redshift 
measurements with the standard $\Lambda$CDM cosmology
and the \texttt{kcorrect} algorithm 
(Blanton et al. 2003) adapted to the AGES data and the NDWFS $B_W$, $I$, and $R$ photometric filters. 
Figure~\ref{hspecRabsz} shows the distribution 
of the galaxies as a function of redshift and absolute magnitude.  The median redshift of the galaxies
is $\langle z\rangle \simeq 0.25$ with an appreciable tail extending to $z\approx 0.6$.  The
change in the density of points at fainter magnitudes is due to the
random 20\% sampling of the fainter galaxies.  To compensate for this effect,
we weight the emission from these galaxies by a factor of 5.  

We also determined the rest-frame $K$-band luminosities, $L_K$, with \texttt{kcorrect}.
We use the $K$-band luminosities to estimate the contribution
of LMXBs to the observed X-ray emission (Eqn.~\ref{eq:LxLK}), and to estimate
the stellar mass 
\begin{equation}
        M_{*} \simeq (8.0 \pm 2.0)\times 10^{10}\left( { L_K \over L_{K*} } \right) M_\odot ,
\label{eq:Mstar}
\end{equation}
based on the results of Colbert et al. (2004, also see Gilfanov 2004).  
To explore how late-type galaxies of different
masses contribute to star formation, we then combine our estimates of
the stellar mass and the star formation rate (determined using Eqn.~\ref{eq:LxSFR})
to compute the Specific Star 
Formation Rate $\rm SSFR = SFR/M_*$ (e.g., Cowie et al. 1996, Guzman et al. 1997, Brinchmann \&
Ellis 2000, Juneau et al. 2004, Feulner et al. 2004, P\'erez-Gonz\'alez et al. 2005).

If, instead of star formation, the X-ray emission is primarily due to accretion onto
AGN, then we can estimate the SMBH growth rates, $\dot{M}_{\rm BH}$, using the scaling of Ba01, 
\beq
\dot{M}_{\rm BH} = 1.76\times 10^{-6}
\left(\frac{\epsilon}{0.1}\right)^{-1}L^{\rm AGN}_{\rm x40}~\rm M_\odot \rm yr^{-1},
\label{eq:Mdot}
\eeq
where $L^{\rm AGN}_{\rm x40}$ is the AGN luminosity in units of $10^{40}$~ergs~s$^{-1}$
and $\epsilon$ is the efficiency of energy 
conversion in AGN. 
Cowie et al. (2003) and Ba05 showed that the X-ray luminosities of both 
obscured and unobscured AGN (with $10^{42}$ erg s$^{-1} < L_{\rm x} <10^{44}$
erg s$^{-1}$) are consistent with a radiative
efficiency of $\epsilon \simeq 0.1$ all the way to $z=0$.
Moreover, the results of AGN models do not agree with observations
if one assumes that lower radiative efficiency AGN dominate
black hole growth, (e.g., Shankar et al. 2004,
Hopkins et al. 2007).
We will therefore
adopt $\epsilon = 0.1$ throughout the paper.

To fit the X-ray luminosities of the AGES galaxies as a function of 
K-band luminosity and redshift, we assume
\beq
L_{\rm x} \propto L_K^\alpha (1+z)^\beta.
\label{double_power_law} 
\eeq
In order to estimate the mean star formation density 
$\dot{\rho}_*$ and nuclear accretion rate density $\dot{\rho}_{\rm BH}$ we must 
extrapolate our results from the magnitude-limited AGES galaxy samples to fainter
fluxes.  By assuming our power-law model (Eqn.~\ref{double_power_law}) holds for fainter sources
and combining the K-band luminosity functions for late- and early-type
galaxies measured by Kochanek et al. (2001), with a change of variables from $L_K$
to $L_{\rm x}$, we compute the X-ray luminosity density as a function of redshift.
The X-ray luminosity density can then be converted to star formation or 
nuclear accretion rate densities using Eqns. (1) or (4), respectively.  

We restrict our analysis to a sample of 6146 ``normal'' galaxies with 
no optical or X-ray evidence of AGN activity.  When stacked, these galaxies produce a total of 486/669 hard/soft 
counts above the background level.  In Table 1, we compare the emission from the normal galaxies
to two classes of sources we exclude from the analysis.  
We eliminated 47 spectroscopically identified AGN (3 broad line and 44 narrow line) that were
\it{not}~\rm identified in the XB\"ootes Survey (Kenter et al. 2005), but these produced a
stacked flux of only 32/27 hard/soft counts.
Much more importantly, we excluded 58 galaxies 
that were identified both by XB\"ootes as X-ray sources and by AGES
as spectroscopic broad line (24) or narrow line (34) AGN.  
Although these 58 AGN represent less than 1\% of our sample by number, they generate 787/1196 hard/soft counts -- 
nearly twice the X-ray flux of the normal galaxies --
and would therefore dominate our results if they were not eliminated from
the analysis (see Fig.~\ref{allLxLKMwinAGNtest}).  Note, however, how similar the composite X-ray spectrum
of the remaining ``normal'' galaxies ($\Gamma _{\rm Normal} \simeq 1.3$)
is to that of these AGN ($\Gamma _{\rm AGN} \simeq 1.3$) and
to that of typical AGN, which account for the bulk of 
Cosmic X-ray Background (CXB; $\Gamma _{\rm CXB} \simeq 1.4$, e.g., Tozzi et al. 2001; Nandra et al. 2004).  
The remaining ``normal'' galaxies therefore provide a representative sampling of the CXB source
population, and, in fact, comprise $\approx 8.8\%$ of the total CXB emission,
based on the soft and hard band CXB flux estimates of De Luca \& Molendi (2004).

%
%FIG1: STRIPS
%
\bfignoc
\centerline{\epsfxsize=9.5cm \epsfbox{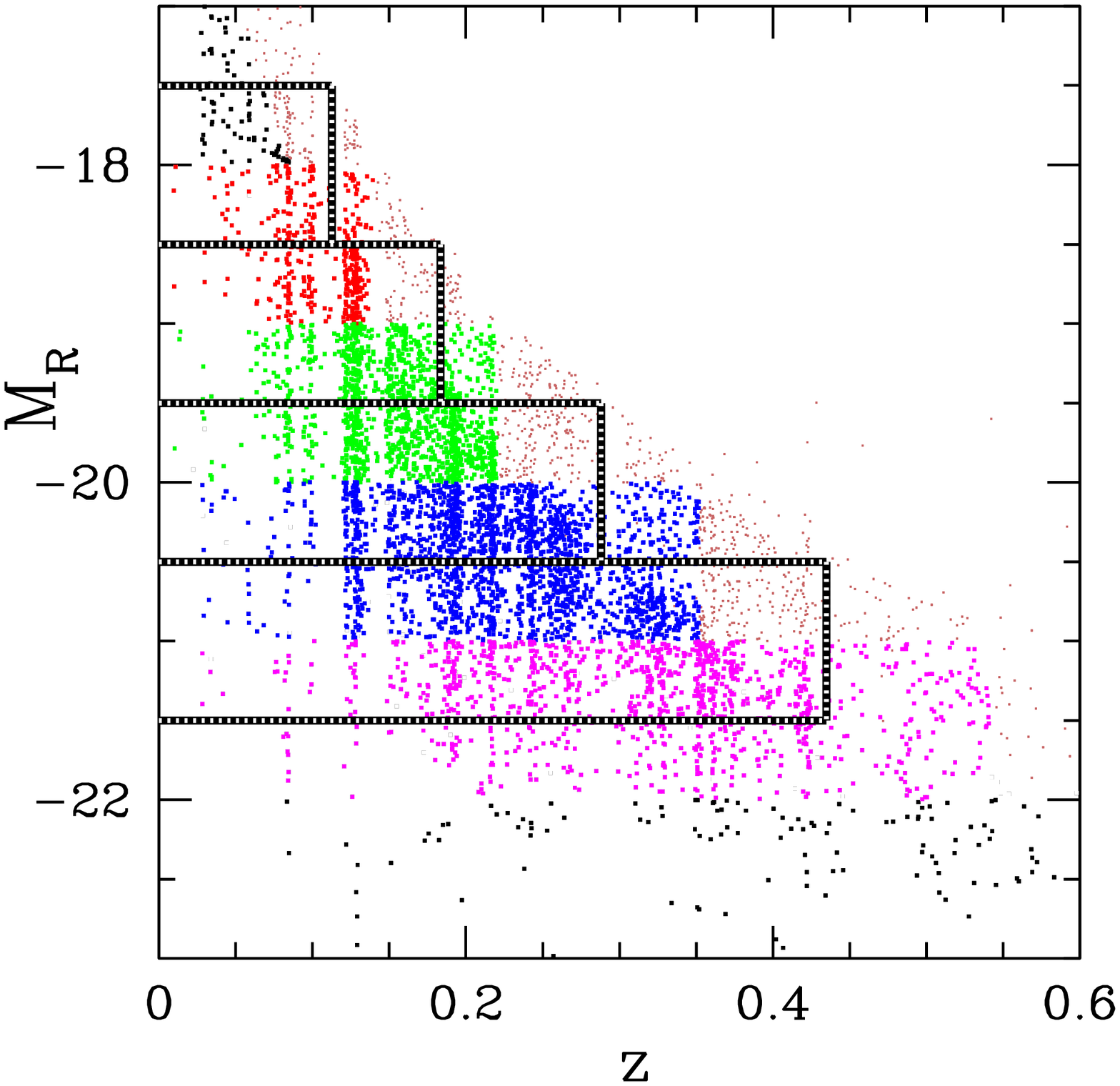}}
\caption{The absolute $R$-band magnitudes of the AGES main sample of galaxies as a
function of redshift. The transition from higher to lower density background points
reflects differences in sampling: all $R < 19.2$ galaxies are present
but only a random 20\% of galaxies with $19.2 < R < 20$ have been sampled.
Four of the eight absolute magnitude strips of galaxies we analyze are outlined.} 
\label{hspecRabsz}
\efignoc
%
%FIG2: STACK
%
\bfignoc
\centerline{\epsfxsize=9.5cm \epsfbox{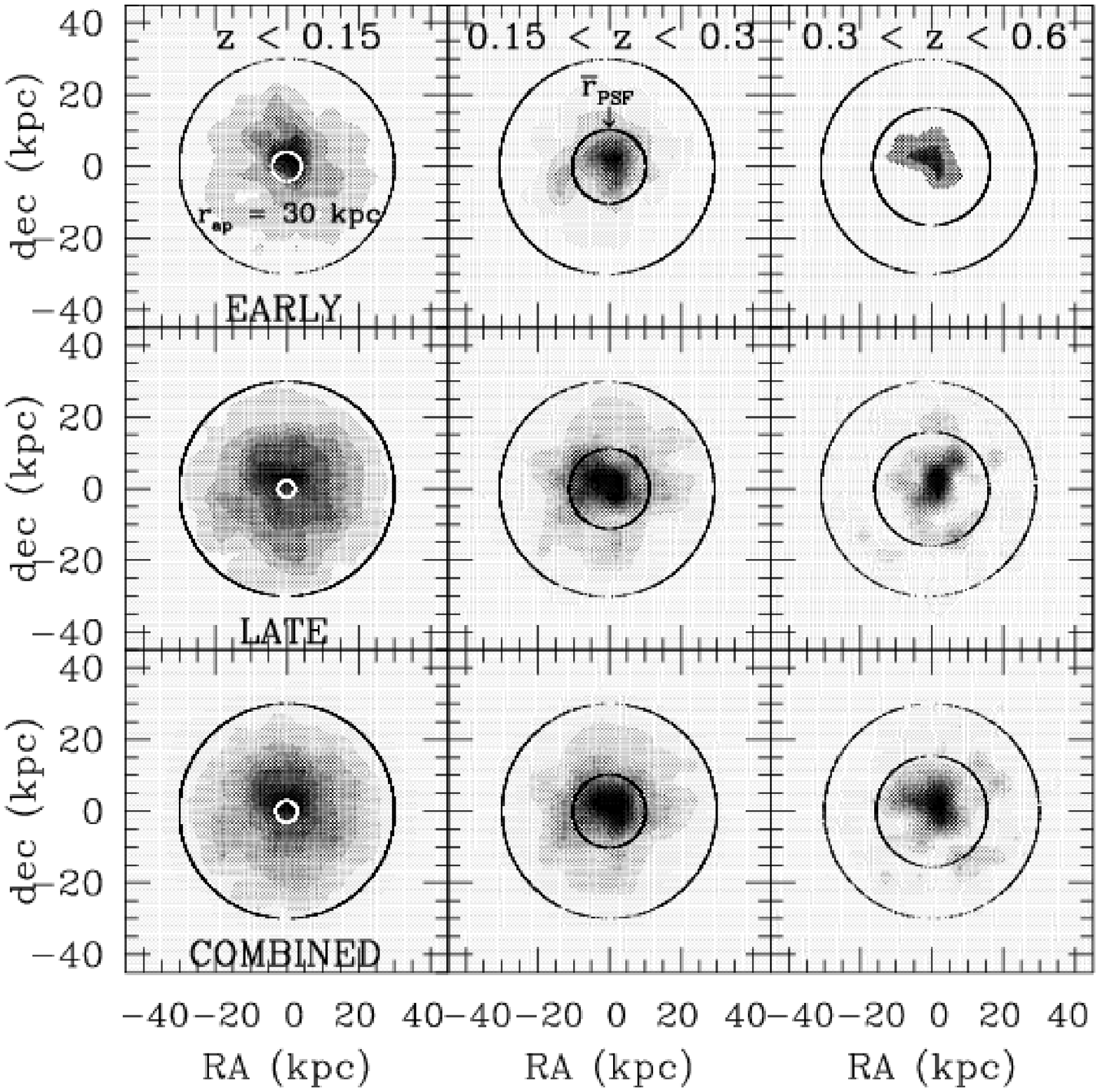}}
\caption{ Total band X-ray images of the stacked early-type, late-type, and combined samples of 
galaxies have been Gaussian-smoothed on a scale of 5 kpc and divided into three redshift bins: $0<z<0.15$, 
$0.15<z<0.3$, and $0.3<z<0.6$.  We have included only galaxies (extended sources) that
had no obvious optical signatures of AGN activity.
In each panel, the outer circle shows the fixed 30~kpc aperture 
we use to extract the signal
and the inner circle shows the size of the 50\%
enclosed energy radius averaged over the field and converted to a physical scale based on the mean redshift of each bin ($\bar{r}_{\rm PSF} \simeq 2,~10$, and 15 kpc, respectively).
}
\label{hspecstack}
\efignoc   
%
%FIG3: NULL
%
\bfignoc
\centerline{\epsfxsize=9.5cm \epsfbox{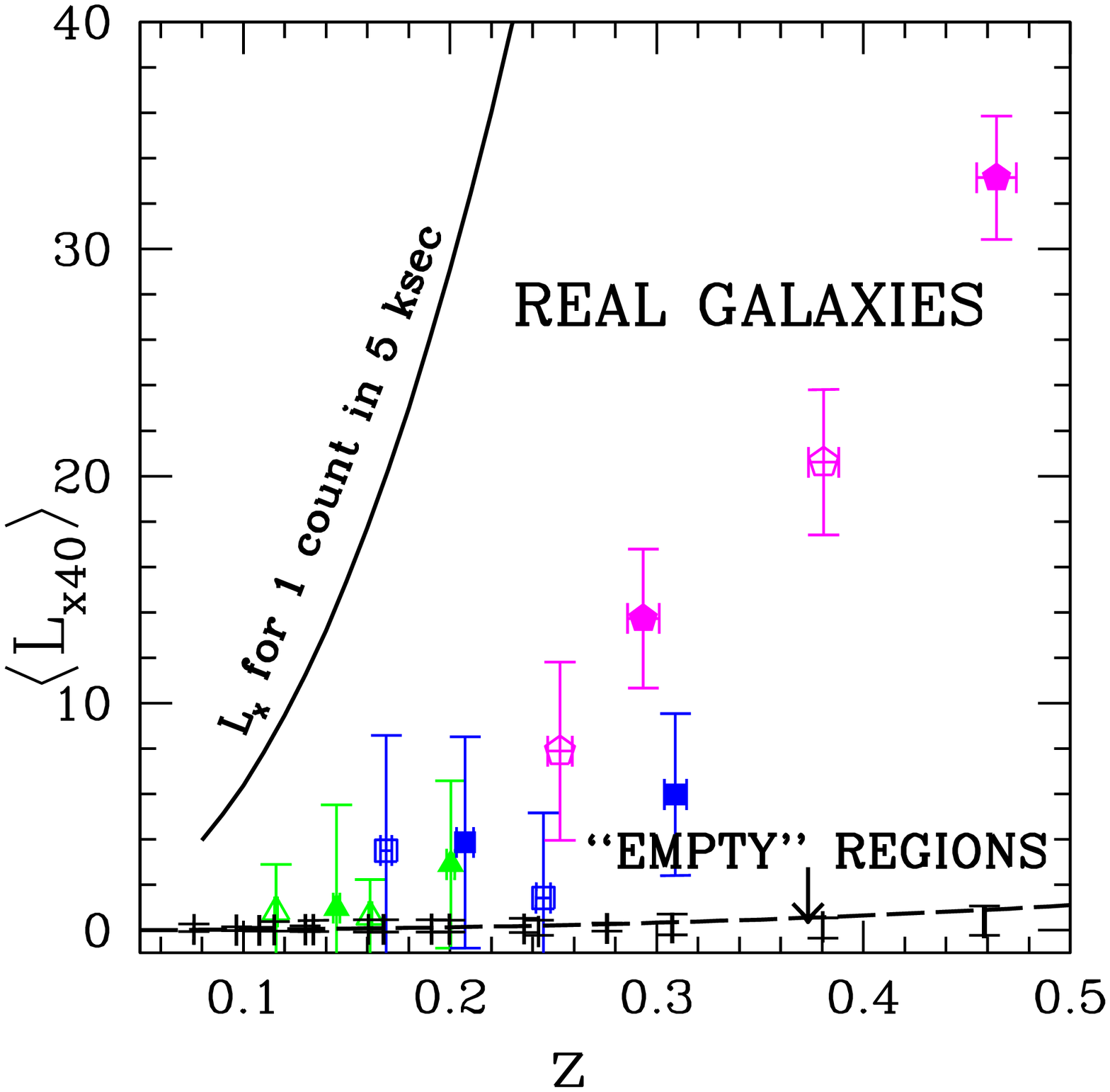}}
\caption{A comparison of the mean X-ray luminosities of the AGES normal galaxies and of
random locations in the XBo\"otes field.
The mean X-ray luminosities (in units of $10^{40}\rm{ergs~s}^{-1}$) of galaxies drawn from
bins in the half-integer (integer)
absolute magnitude strips with $M_R < -19$ (see Figure
\ref{hspecRabsz}) are designated, from faint to bright, by open (closed) 
triangles, squares and pentagons. 
No statistically significant X-ray flux was detected 
from the intrinsically faintest galaxies. 
Most of the trend here is
due to the higher redshifts of the more massive (and luminous) galaxies in our flux-limited sample.
The dashed line shows our expected sensitivity limit given the 5~ksec exposure time
and the typical number of objects in each bin (see \S3). 
Note that through stacking we are able to detect average X-ray signals
that are much less than the luminosity equivalent of 
one total band count (solid line).
}
\label{randhspecLx}
\efignoc

\section{Method, Monte Carlo Simulations and Tests}

The basic principle of a stacking analysis is simple.  The X-ray images of a large number of
sources are stacked to determine the mean properties of a group of objects
that are individually undetectable (Brandt et al. 2001; Nandra et al. 2002; Hornschemeier
et al. 2002, Georgakakis et al. 2003, Lehmer et al. 2005, Laird et al. 2005, Laird et al. 2006, Lehme et al. 2007).
Quantitatively, if stacking a sample
of $n$ objects yields $N_s$ counts in a signal aperture of area $A_s$ and
$N_b$ counts in a background annulus of area $A_b$, then the mean number of source
counts per object in the signal region is
\begin{equation}
      \langle N \rangle_{\rm src} = { 1 \over n } \left[ N_s - { A_s \over A_b} N_b \right].
\label{Nsrc}
\end{equation}
The uncertainties in $\langle N \rangle_{\rm src}$ can be computed either
by assuming a Poisson distribution of X-ray counts per object,
\begin{equation}
     \sigma = { 1 \over n } \left[ N_s + { A_s^2 \over A_b^2} N_b \right]^{1/2},
\end{equation}
or by bootstrap resampling of both the galaxy and X-ray photon catalogs (event lists).
Adopting the latter method, we generate 100 bootstrap
samples by randomly drawing new lists of galaxies and X-ray photons (with
replacement) from the input catalogs until each
bootstrap sample contains the same number of galaxies and X-ray photons
as the real catalogs.
Each mock data set is then subjected to the same analysis as the real data.
Our error estimates are defined by the range encompassing 68\% of the bootstrap results
about the true mean.
This approach will produce more realistic error bars
than the Poisson statistics of the stacked image, particularly
if the net flux is dominated by a small fraction of
the objects included in the analysis.

As a general proof of principle for the subsequent analyses, Fig.~\ref{hspecstack} 
shows images of the stacked early-type, late-type and combined samples of galaxies in three 
redshift bins, excluding direct X-ray detections and galaxies with optical signatures of AGN activity.
Based on the Monte Carlo simulations we discuss in 
the next section, we adopt a 30~kpc signal aperture and a 30 to 60~kpc
background annulus about the center of each galaxy.  
Each panel shows this signal aperture and the field-averaged point spread function
(PSF), i.e., the 50\%
enclosed energy radius averaged over the field and converted to a physical scale based
on the mean redshift of each bin.
We clearly detect a signal from these ``normal'' galaxies. 

To understand the X-ray evolution of the galaxies quantitatively, we bin them
by absolute magnitude, spectroscopic type and redshift.  We analyze the galaxies in 
staggered, 1~mag wide absolute magnitude strips (see Fig.~\ref{hspecRabsz}),
beginning at integer and half-integer
absolute magnitudes (i.e., $-17.5 > M_R > -18.5$, $-18 > M_R > -19$,
$-18.5 > M_R > -19.5,~ .~ .~ .~ -20 > M_R > -21$).
We divide each strip into 
into two redshift bins that contain roughly equal numbers of galaxies
after correcting for the sparse sampling.  
We also analyze the combined, late-type, and early-type galaxy samples 
separately.
For each bin we then determine the mean X-ray and rest-frame
$K$-band luminosities of the member galaxies.  
Because the X-ray and optical luminosities of galaxies are approximately linearly correlated 
(e.g., David, Jones, \& Forman 1992; Shapley et al. 2001), 
the X-ray to optical luminosity ratio should provide a measure of the 
redshift evolution of normal galaxy X-ray emission that is essentially independent
of galaxy (optical) luminosity and stellar mass (Eqn.~\ref{eq:Mstar}).

\bigskip
\bigskip
\bigskip

In addition to spectroscopically identified AGN, we excluded directly detected
X-ray sources that did not satisfy 
\beq
L_{\rm x} \leq 6.6\times 10^{42} \rm{erg~s^{-1}}~\rm{(luminosity~limit)},
\label{LumLimit}
\eeq
which corresponds to two total band counts observed at the
upper redshift limit for the normal galaxies: $z=0.6$.  The most luminous detected sources at lower redshifts,
with $10^{41}$ erg s$^{-1} < L_{\rm x} <10^{42}$ erg s$^{-1}$,
are within the range possible for star forming galaxies, (e.g., Norman et al. 2004).
Moreover, our bootstrap uncertainty estimates are close
to the Poisson limits, which means that the X-ray flux
cannot be dominated by a small number of luminous sources.

As an additional check on our ability to detect galaxies and to diagnose problems
with background subtraction or contamination, we constructed
a null sample to compare with the fluxes of the real sources.
To create the null sample we kept the redshifts and 
(optical) luminosities of the galaxies fixed but assigned them random positions in the field,
excluding locations closer than 60~kpc to an AGES galaxy.  We then analyzed 
the random catalog in the same manner as the actual data.  In Fig.~\ref{randhspecLx}, 
we compare the luminosities of the real galaxies and the ``empty'' regions.
While the real galaxies generate a significant signal
in all but the faintest ($M_R \gtorder -18$) bins, the mean luminosities 
of the random fields are always consistent with zero.
The uncertainties
in the random signal closely follow the expected trend of
$\Phi_{\rm min} 4\pi d^{2}_{L}(z)$, where 
$\Phi_{\rm min} \approx 4\times 10^{-17}$~ergs~s$^{-1}$~cm$^{-2}$ is the sensitivity
of the observations divided by the square root of the number of galaxies in a typical
bin.  Stacking the galaxies in our luminosity and redshift bins 
generally provides $\gtorder 25$ times the sensitivity associated with the
5 ksec exposure time for individual sources.  Over the range of redshifts we probe, 
this flux limit corresponds to threshold luminosities of
$\approx 10^{38}$~ergs s$^{-1}$ to $10^{40}$~ergs s$^{-1}$.

The X-ray properties of our stacked galaxy samples are
summarized in Tables 2 and 3. Table 2 provides upper limits on the hard and soft band
X-ray luminosities of the optically faintest AGES galaxies that were either not detected
in X-rays or had net stacked emission
with S/N $< 1$ in all X-ray bands.
Table 3 provides the data for the detected (S/N $> 1$) galaxies.

\subsection{Monte Carlo Simulations}

We used Monte Carlo simulations to test our analysis methods and to
refine our choices of parameters.  We started by generating a mock galaxy
catalog using the Brown et al.~(2001) $R$-band luminosity function,
a simple model for galaxy evolution, and the AGES main galaxy
sample selection criteria including the sparse sampling.  The general properties of the resulting
synthetic catalog are quite similar to those of the real one. The absolute
magnitude strips are a case in point. From faint to bright magnitudes,
the real strips with integral, absolute magnitude boundaries in Fig.~\ref{hspecRabsz}
contain 471, 1621, 2526 and 1496
galaxies including the corrections for sparse sampling, while the corresponding
strips in the synthetic catalog contain 614, 1630, 2218 and 1527 galaxies.

To simulate the X-ray data, we began by distributing $\approx$ 350,000 background photons at
random positions in the field with surface densities of $1.9 \times 10^{-3}$ hard (2--7
keV) and $0.9 \times 10^{-3}$ soft (0.5--2 keV) photons arcsec$^{-2}$, based on the
background estimates for the XBo\"otes survey (Kenter et al. 2005). We then assigned
each galaxy an extended X-ray component and a point-like nucleus.  For simplicity, the
extended emission consists solely of soft photons and the AGN component consists
solely of hard photons. These components have mean luminosities of
\beqa
L_{\rm x40,soft}= 5(L/L_{*})(1+z)^3 \nonumber \\
L_{\rm x40,hard} = 5(L/L_{*})(1+z)^4,
\label{eq:Input_Lum}
\eeqa
respectively, in units of $10^{40}$~ergs~s$^{-1}$, where $L/L_*$ is the rest-frame ($R$-band) optical
luminosity in units of the characteristic luminosity, $L_*$, at the knee of the Schecter luminosity
function.  The redshift scaling and the correlation between the X-ray and optical
luminosities are motivated both by previous studies
(e.g., David, Jones, \& Forman 1992; Shapley et al. 2001, Brand et al. 2005) and by our own results (see \S4).

Based on the X-ray spectrum of the AGES galaxies, we converted the X-ray fluxes to counts
using soft and hard band flux limits of $1.05 \times 10^{-15}$~ergs~cm$^{-2}$~s$^{-1}$
and $3.65 \times 10^{-15}$~ergs~cm$^{-2}$~s$^{-1}$, respectively
and K-corrections of $(1+z)^{(\Gamma-2)}$, with $\Gamma = 1.29$ (see \S2.1).
The hard emission was distributed about
the galactic centers using a Gaussian model of the CXO PSF with a 50\%
enclosed energy radius of
\beq
r_{\rm PSF} = 0\farcs5 + 6\farcs0\left(\frac{D}{10'}\right)^{2},
\label{eq:rpsf}
\eeq
where $D$ is the randomly assigned distance of the galaxy from the CXO pointing center
in arcminutes
(see the CXO Proposers' Guide\footnote{http://cxc.harvard.edu/proposer/POG/html}).
We modeled the extended emission as a Gaussian of dispersion $r_L=10(L/L_*)^{1/2}$~kpc
convolved with the PSF to get an overall radial profile of
\beq
   P_{\rm extended}(R) \propto {\rm exp}\left[\frac{-R^2}{2(r^{2}_{L} + r^{2}_{\rm PSF})}\right].
\label{eq:Pr}
\eeq
In this model the X-ray surface brightness of the galaxies is independent of optical luminosity.
%
% FIG 4: SIM PROFILE 
% 
\bfignoc
\centerline{\epsfxsize=9.5cm \epsfbox{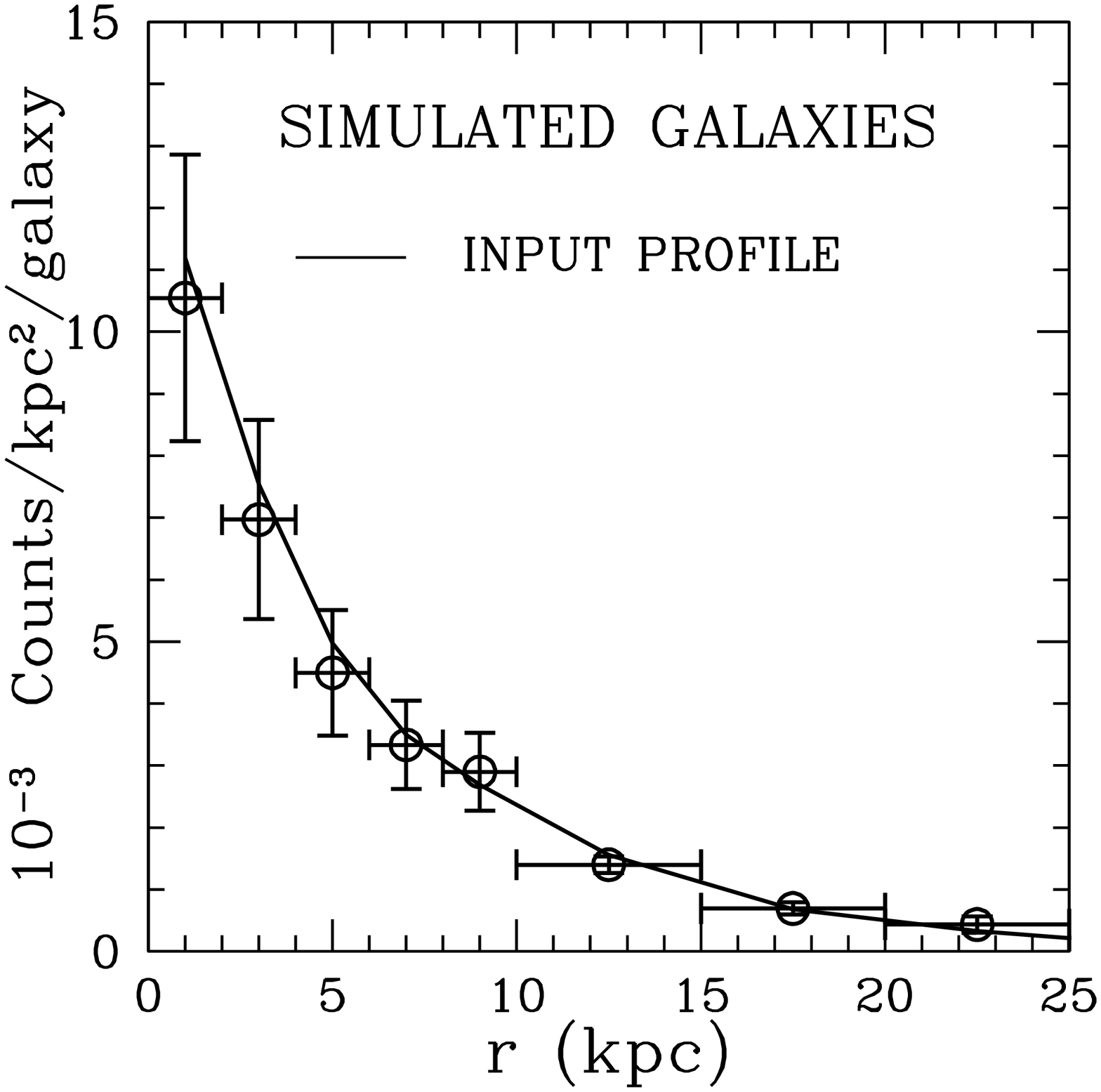}}
\caption {
Monte Carlo test of the stacking procedure.  The solid curve
shows the input total band emission profile in our Monte
Carlo simulation, and the open circles show the
profile extracted by the stacking analysis.  The extracted
profile agrees with the input profile in all radial bins.
}
\label{Nvrhspecsimzpg}
\efignoc

%
%FIG5: L_x/L_K for AGN vs. Normal Galaxies
%
\bfignoc
\vspace{0.3cm}
\centerline{\epsfxsize=9.2cm \epsfbox{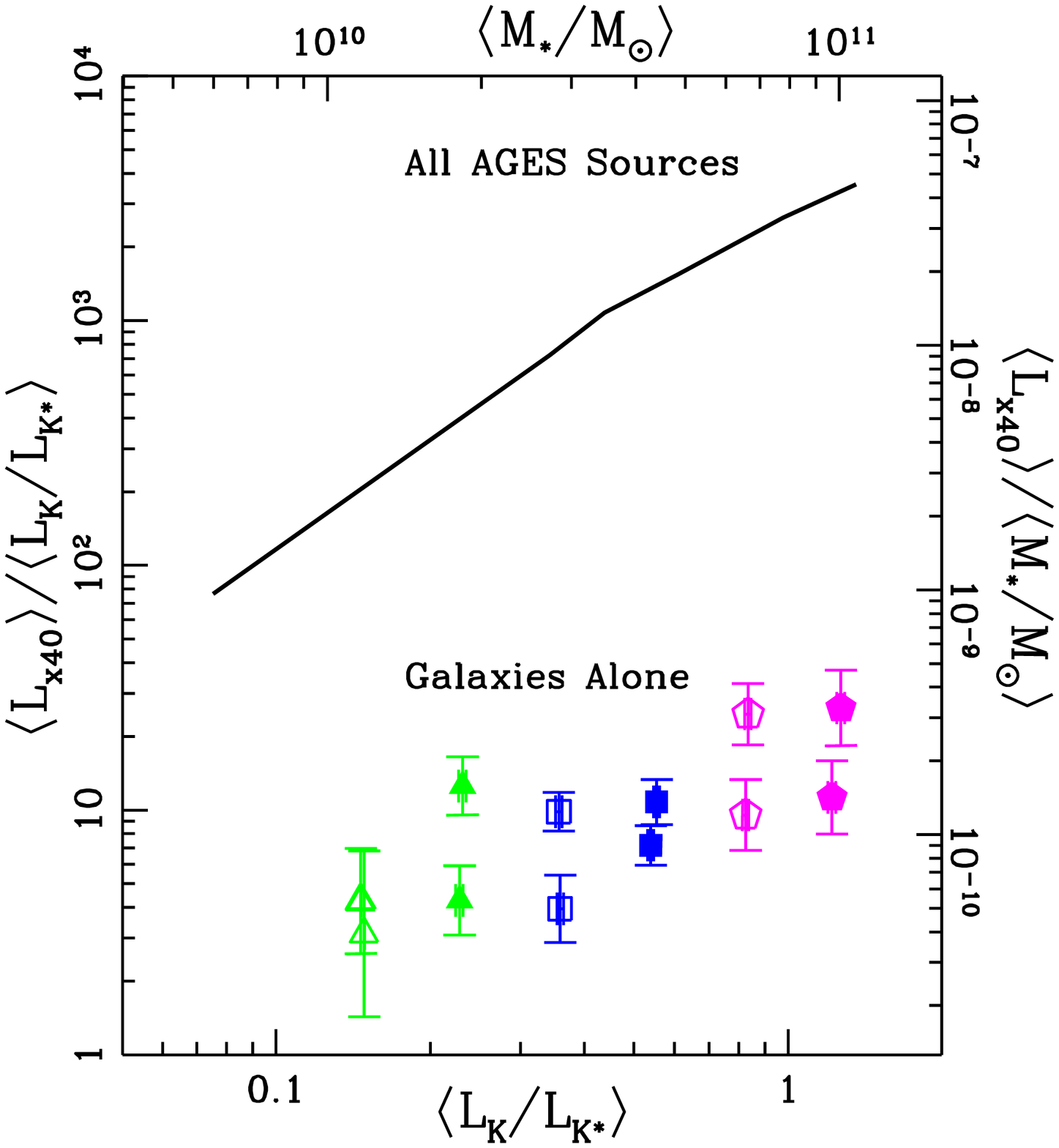}}
\caption{The specific X-ray luminosity
of all AGES sources compared to that of the normal galaxy sample.
The solid line is a fit to the
sum of the mean luminosity of the normal galaxies and the mean luminosity of the
58 X-ray detected
AGN (see \S2.2 and Tables 1 and 3) that lie within the absolute magnitude 
and redshift ranges we consider. 
The binned data shows the 
mean luminosity evolution of the normal galaxies alone; the point type conventions are
the same as those used in Fig.~\ref{randhspecLx}. 
The two points at
each value of $L_K$ demonstrate the evolution from the low to high redshift bin
in each absolute magnitude strip (Fig. 1).
}
\label{allLxLKMwinAGNtest}
\efignoc

\subsection{Radial Emission Profiles}

We first verified that our analysis correctly extracts the X-ray emission profiles
by analyzing the data for the roughly 800 galaxies in the mock catalog with 
$z\leq 0.1$ (mean redshift $\langle z\rangle\simeq 0.07$) and within $10'$ of the pointing center.  
We consider only these lower redshift sources
because, in spite of the CXO's high, on-axis resolution, the
field-averaged PSF still approaches 2~kpc at $z\simeq 0.07$.
Fig.~\ref{Nvrhspecsimzpg}
shows the extracted profile
for the total band.
Using a 30 kpc signal aperture and a 30--60 kpc background subtraction annulus, we recover
the input profile (and background level)
to within the 1$\sigma$ bootstrap uncertainties.
We carry out a similar analysis on the real data in \S4.1.  

%
%FIG6   PROFILES   
%
\bfignoc
\centerline{\epsfxsize=9.5cm \epsfbox{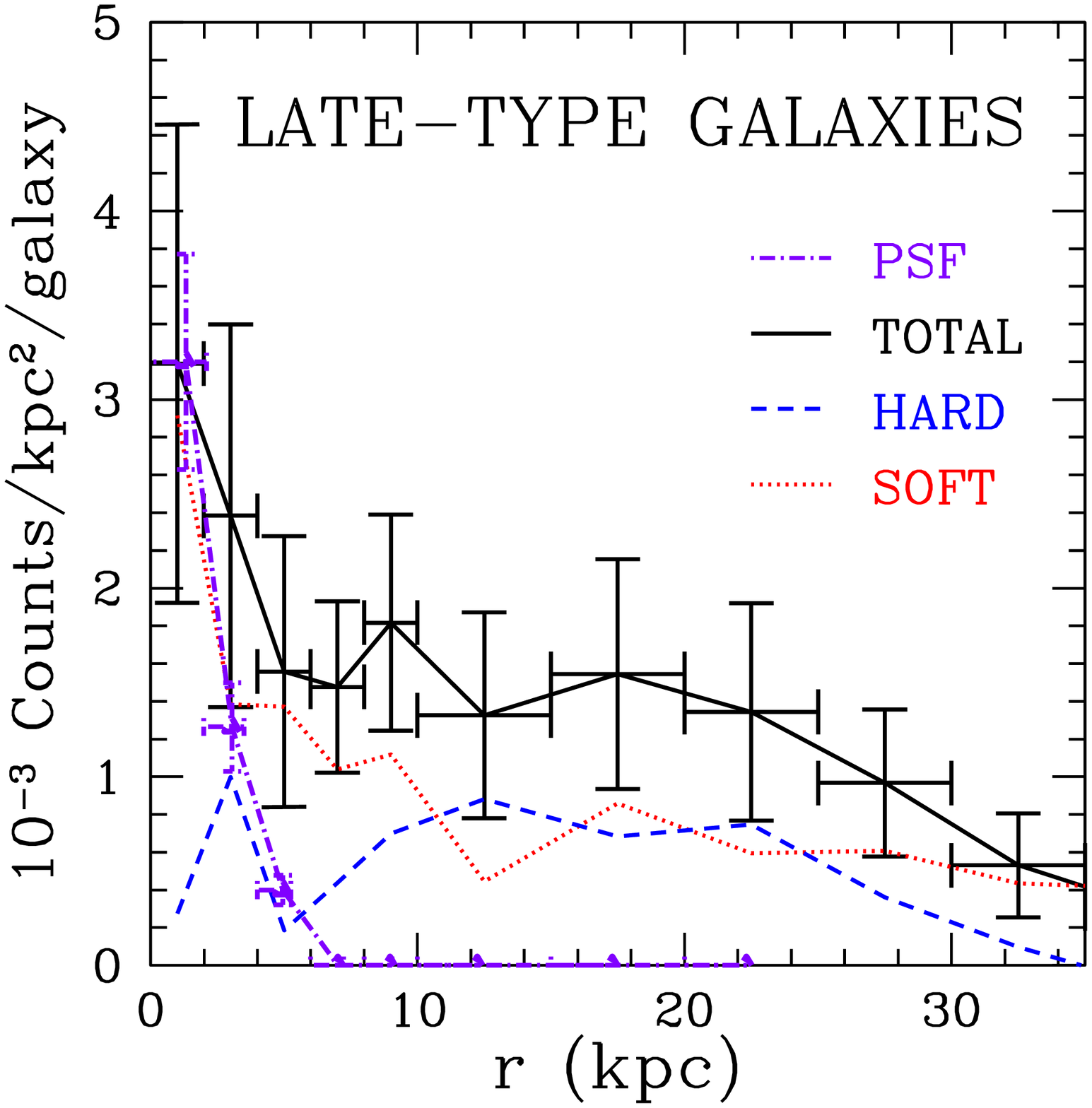}}
\vspace{1.5cm}
\centerline{\epsfxsize=9.5cm \epsfbox{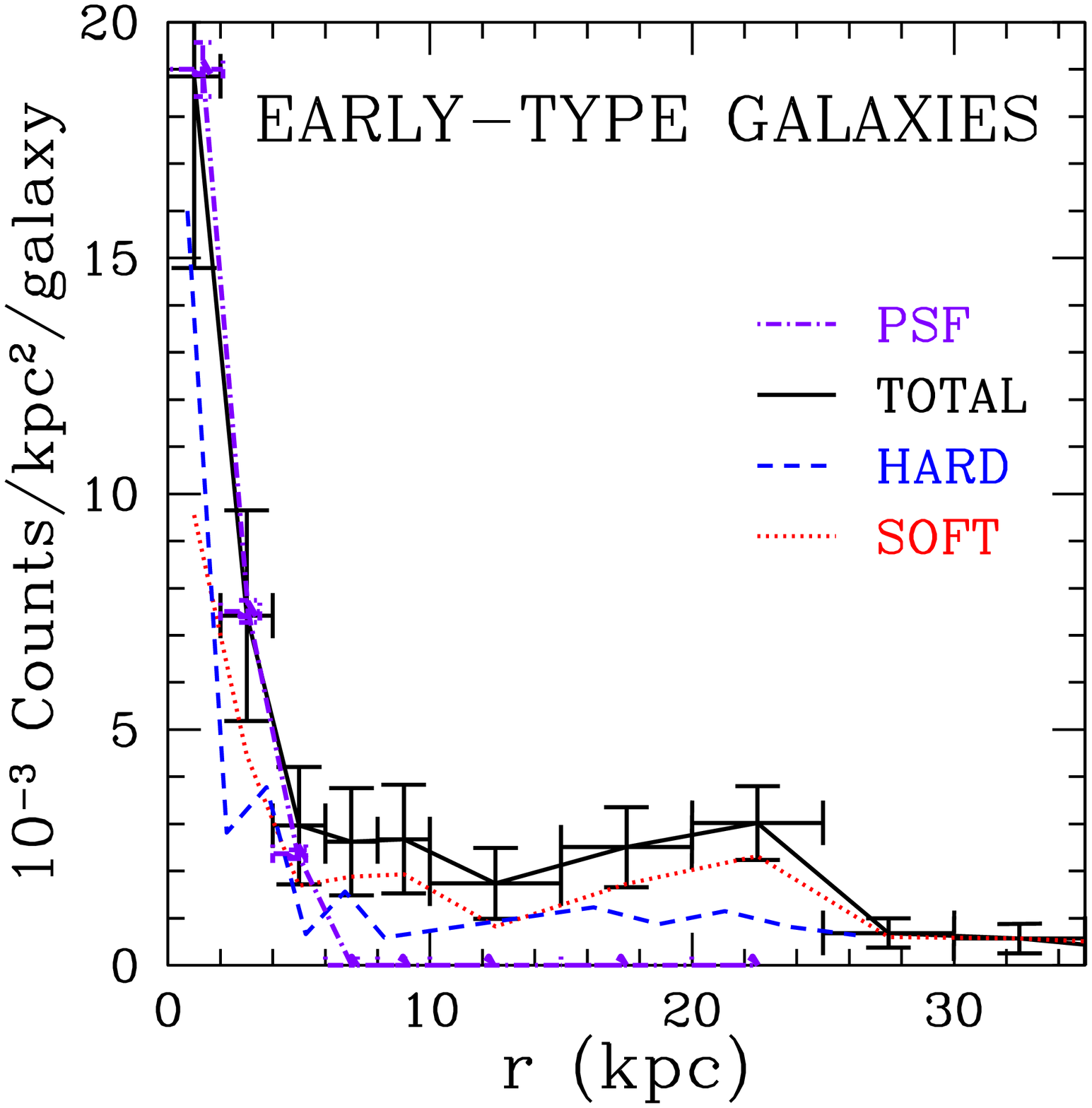}}
\caption{Background-subtracted radial emission profiles of
360 late-type (top) and 128 early-type (bottom) galaxies with $z\leq 0.1$,
a mean redshift of $\langle z\rangle \approx 0.07$
and within $10'$ of CXO field centers.  The solid, dashed, and dotted curves represent the
total, hard, and soft band profiles, respectively, in each panel.
The dot-dashed curve shows an empirical model of the PSF, based on the radial
emission of $\approx 500$, $z\gtorder 1$ AGN within $10'$ of a CXO field center.
We have shifted the normalization of the PSF model to match the amplitude of the
galaxy profile in each panel.}
\label{Nvrhspec}
\efignoc
%
%FIG7: HR   
%
\bfignoc 
\centerline{\epsfxsize=9.5cm \epsfbox{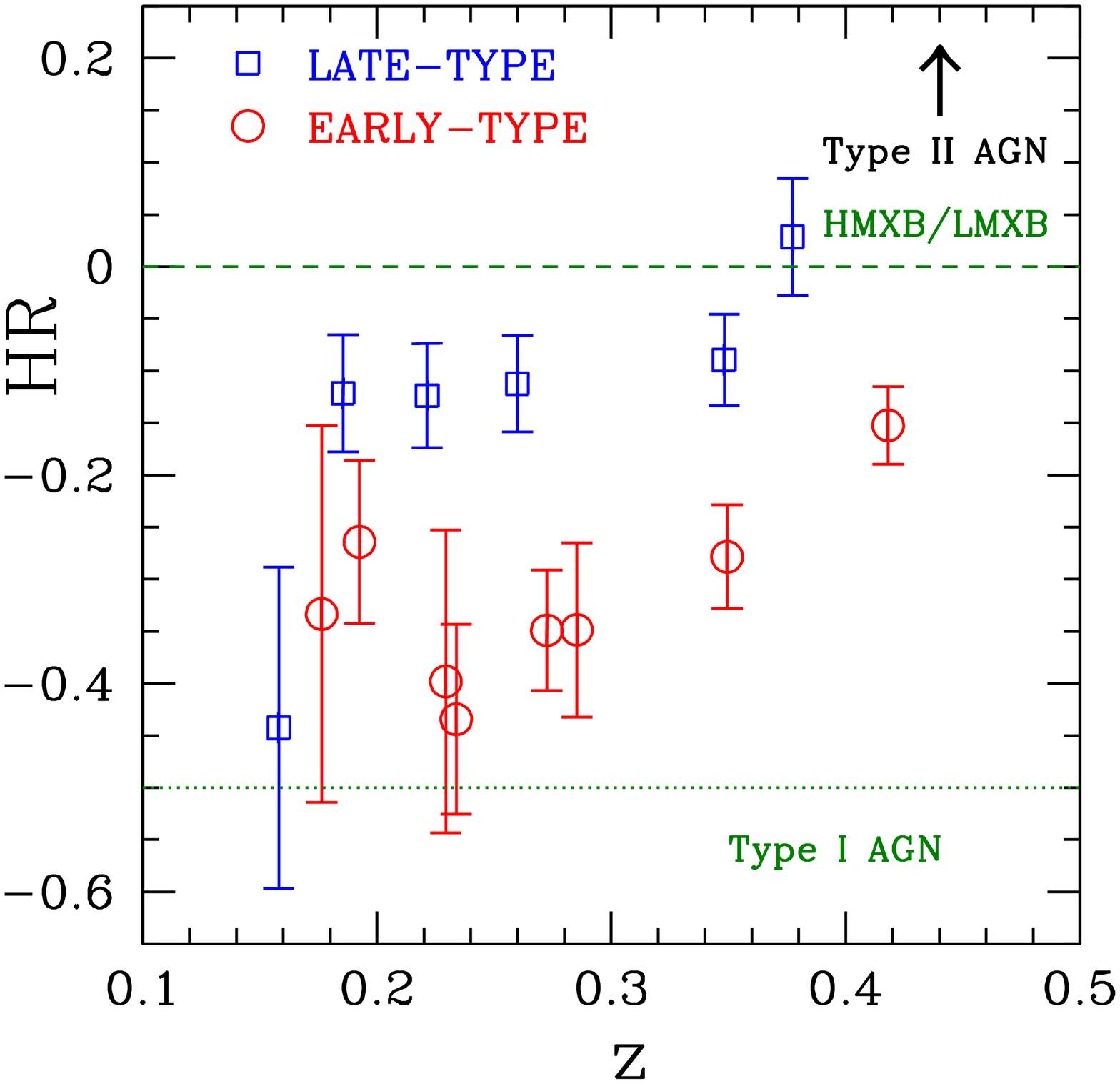}}
\caption{Redshift evolution of the Hardness Ratio (HR) for the late-type (squares)
and early-type (circles) samples of galaxies. 
The late-type galaxies exhibit slightly harder spectra than the
early-type galaxies, which is most likely due to the early-types'
characteristically larger halos of
soft X-ray emitting gas. 
Typical HRs of HMXBs and LMXBs (dashed line) as well as Type I (dotted line)
and Type II AGN (typically above the dashed line)
have been included for reference.  
}
 \label{HR}
\efignoc

At higher redshifts, where we cannot spatially resolve the emission, we concentrate on extracting
the redshift evolution of the X-ray luminosity.  This involves
a trade-off between the signal-to-noise ratio (S/N) and the need for aperture corrections.
Larger apertures capture more signal but introduce extra noise from the
background, while small apertures lose some of the signal and require
more substantial aperture corrections. 
Redshift-dependent biases will
be minimized by using a fixed angular aperture for point sources and a (sufficiently large) fixed 
physical aperture for extended sources.
Since we expect a significant contribution from
extended emission in the real data (e.g.,  Fig.~2), we focused on fixed physical apertures with proper
radii of  $R_{\rm ap}=$ 10, 20, 30, 40, and 50 kpc and background
annuli extending over the larger of 30--60 kpc or $R_{\rm ap}$--$2R_{\rm ap}$. 
Adopting a signal aperture of $R_{\rm ap}=30$ kpc (at the mean redshift of each bin)
led to the most accurate
reproduction of the input profiles with the highest S/N; our detection criterion is S/N$~>1$.
This fixed physical aperture corresponds to angular
apertures of $\sim 15'' - 5''$  over the range of mean redshifts we analyze ($0.1 < z < 0.5$, see Table 2 and Fig.~2).
The size of the field-averaged PSF begins to exceed that of significantly smaller apertures, 
particularly in the higher redshift bins (see Fig.~2).  For 
point sources we could apply a redshift-dependent aperture correction to remedy this problem,
but it is impossible to do so for extended sources when we are uncertain of their intrinsic profiles.  
For larger apertures,
our signal-to-noise ratio begins to drop due to the
increasing fraction of background photons included in the signal region.
In the Monte Carlo simulations, small deviations ($\ltorder 10$ kpc) 
from $R_{\rm ap}$ = 30 kpc have little effect on the results or the 
signal-to-noise ratio but larger (factor of 2) changes in the aperture lead to serious problems.

Using $R_{\rm ap}=30$~kpc and 30--60 kpc background annuli, which we adopt throughout the remainder of the paper,
we successfully recover the input luminosity
evolution (Eqn.~\ref{eq:Input_Lum}); i.e., 
when we fit our results with a double power law model (Eqn.~\ref{double_power_law}), 
we find scalings for the soft and hard band X-ray luminosities
that are statistically consistent with the input evolution rates and normalization:
\beqa
 L_{\rm x40,soft} = (4.7 \pm 1.7)\left( \frac{L_K}{L_{K*}}\right)^{1.0\pm 0.2}(1+z)^{2.7 \pm 0.8}, \nonumber \\
 L_{\rm x40,hard} = (4.0 \pm 2.4)\left( \frac{L_K}{L_{K*}}\right)^{0.9\pm 0.3}(1+z)^{3.4 \pm 1.1}.
\eeqa

\bigskip
\bigskip

\section{Results}

We now turn to the real sources,
starting with a comparison of the X-ray emission from normal galaxies and from spectroscopic AGN.
Fig.~\ref{allLxLKMwinAGNtest} shows the specific X-ray luminosity, 
$\langle L_{\rm x40}\rangle/\langle L_{K}/L_{K*}\rangle$, 
as a function of $\langle L_{K}/L_{K*}\rangle$ for both the normal galaxies and the full sample including the
X-ray and spectroscopically flagged AGN. 
With the top and right axes, we also show the inferred scaling with stellar mass, based on Eqn.~(\ref{eq:Mstar}), although this
conversion does not apply when the bright AGN dominate the $K$-band flux.
While the AGN have significantly larger X-ray luminosities than the ``normal'' galaxies, the
``normal'' galaxies are clearly detected.  The two points at each $L_K$ show the evolution of the normal galaxy X-ray emission
from the low to high redshift bin
in each absolute magnitude strip (Fig. 1). When we fit the normal galaxy data as a 
double power law in K-band luminosity and redshift (Eqn.~\ref{double_power_law}), 
we find that
\beqa 
       L_{\rm x40}       = (6.0\pm1.0)\left( \frac{L_K}{L_{K*}} \right)^{1.1\pm 0.2} (1+z)^{3.2\pm0.8}, \nonumber \\
       L_{\rm x40, hard} = (4.6\pm0.9)\left( \frac{L_K}{L_{K*}} \right)^{1.2\pm 0.3} (1+z)^{3.5\pm1.0}, \nonumber \\
       L_{\rm x40, soft} = (2.0\pm0.5)\left( \frac{L_K}{L_{K*}} \right)^{1.1\pm 0.2} (1+z)^{2.7\pm0.7}
\eeqa
for the total, hard, and soft X-ray bands, respectively, in units of $10^{40}$~ergs~s$^{-1}$.
Because our results indicate that $L_{\rm x}$ is roughly proportional to $L_K$ in all X-ray bands, 
the specific X-ray luminosities of the galaxies
should evolve similarly regardless of their optical luminosity or mass.  

The X-ray luminosities we measure are roughly an order of magnitude brighter than the contribution expected from
LMXBs (Eqn.~\ref{eq:LxLK}), so we are left with three major candidates for the origin of the emission -- 
HMXBs associated with recent star formation, accretion onto supermassive black holes (AGN), and hot gas.  
We use three tests to try to determine the dominant source(s). 
We first use fits to the AGES spectra to 
divide the normal galaxies into late-type and early-type subsamples.
Doing so allows us to
compare the X-ray emission from galaxies with and without (optical)
spectroscopic signatures of star formation.
The resulting subsamples contain 3178 late-type galaxies and 2968 early-type galaxies. 
Second, we examine the radial emission profiles of the low redshift galaxies, since emission 
due to star formation and hot gas will be more spatially extended than emission from AGN. 
Third, we compare the hardness ratios of the galaxies, in terms of the
hard and soft band counts $C_{\rm hard}$ and $C_{\rm soft}$,
$HR=(C_{\rm hard}-C_{\rm soft})/(C_{\rm hard}+C_{\rm soft})$, 
to the typical values expected for X-ray binaries, AGN and hot gas.

\subsection{Radial Emission Profiles}

Fig.~\ref{Nvrhspec} shows the average, background-subtracted soft, hard, and total band emission profiles of 488
(360 late-type and 128 early-type)
low redshift ($z<0.1$),
galaxies within $10'$ of the CXO optical axis,  
excluding X-ray sources
that violate our luminosity limit (Eqn.~\ref{LumLimit}) and AGN identified
through optical spectroscopy. 
The mean redshift of these galaxies is $\langle z\rangle \simeq 0.07$.  
Rather than use a model for the PSF (Eqn.~\ref{eq:rpsf}), we determined it
empirically by averaging the emission profiles of $\approx$ 500 $z>1$ X-ray
selected AGN from the XBo\"otes survey.  As with the galaxies,
we used only AGN sources
within $10'$ of the Chandra field centers and stacked the emission based on the
optical positions. We then applied the angular corrections needed to put the PSF model on the
same physical scale as the galaxies.  

We find that the emission profiles of the late-type galaxies are clearly extended while those of the
early-type galaxies have a more pronounced nuclear component.
We use the Kolmogorov-Smirnov (K-S) test to compare the radial distribution of (total band)
photons from the late-type galaxies, early-type galaxies, and high redshift AGN. 
Within 5 kpc of the stacking centers, we find that the early-type galaxy distribution ($\propto r^{-1.1}$)
is consistent with the AGN distribution (K-S likelihood $97\%$),
while the late-type galaxy distribution ($\propto r^{-0.4}$) is not (K-S likelihood $3\%$).
However, if we analyze the full extraction region, neither galaxy profile is consistent with the AGN model
(K-S likelihood $<~1\%$).
Between 10 and 30 kpc, both galaxy types
show significant ($\gtorder 2.5~\sigma$) emission, which we attribute
primarily to diffuse gas and LMXBs. Soft X-ray emitting gas
has been observed out to $\gtorder 20$ kpc around late-type (e.g., Strickland et al. 2004) and
early-type (e.g., Mathews \& Brighenti 2003) galaxies, as have LMXBs
(late-type: e.g., Wang et al. 1999; early-type: e.g., Kim et al. 2006), leading to extended profiles like those in Fig. 6.
Our results imply that the emission from the low-redshift early-type galaxies
comes from a combination of AGN, hot gas, and LMXBs.  In contrast, because
the emission from the late-type galaxies is significantly less centrally concentrated than the emission of AGN -- 
even within a radius of 5 kpc -- the dominant source of the late-type galaxy flux
is most likely a radially extended distribution of HMXBs and LMXBs. 

\subsection{Hardness Ratios}

In Fig.~\ref{HR} we show the hardness ratios (HRs) based on the hard and soft band counts
of the two galaxy classes.  To improve the statistical significance of our HRs, we combine hard and soft counts
from bins of comparable redshifts in adjacent absolute magnitude strips (Fig.~1).  For reference, we compare these
HRs to those observed in these bands for 
accreting binaries (HR $\gtorder 0$, e.g.,  Muno et al. 2004),
Type I AGN (HR $\simeq -0.5$, e.g., Rosati et al.  2003; Franceschini, et al. 2005), 
Type II AGN ($0 \leq \hbox{HR} \leq 1$, e.g., Rosati et al. 2003), and hot gas with a temperature of
0.5--1~keV ($\hbox{HR} \simeq -1$). 

In general, we observe that the late-type galaxies have 
slightly harder spectra than the early-type galaxies and that the spectra of both galaxy types
become somewhat harder with increasing redshift. This is
most likely a consequence of the larger amount of soft X-ray emitting gas in earlier type
galaxies and the steadily increasing contributions of HMXBs and AGN, respectively, 
to the late-type and early-type galaxy spectra at higher redshifts.

\subsection{Luminosity Evolution}

To isolate the dependence of $L_{\rm x}$ on $L_{K}$, we examined the
evolution of the X-ray luminosity of the early and late-type galaxies
in three narrow redshift slices: $\langle z \rangle = 0.20 \pm 0.02$, $\langle z \rangle = 0.25 \pm 0.02$,
and $\langle z \rangle = 0.35 \pm 0.03$.  As shown in Fig.~\ref{fixedzLxLK}, the specific X-ray luminosities are
relatively flat in $L_K$, indicating an approximately linear correlation between $L_{\rm x}$ and $L_K$
as found previously (e.g., David, Jones, \& Forman 1992; Shapley et al. 2001). 
If we fit these results with a power law model, $L_{\rm x} \propto L_K^\alpha$, 
we find consistent values of $\alpha$ 
among the three redshift slices for both the late-type
($\alpha^{\rm LATE}_{\rm hard} = 1.19 \pm 0.31$; $\alpha^{\rm LATE}_{\rm soft} = 1.12 \pm 0.15$)
and early-type
($\alpha^{\rm EARLY}_{\rm hard} = 1.09 \pm 0.23$; $\alpha^{\rm EARLY}_{\rm soft} = 1.07 \pm 0.15$) galaxies.

Figs.~\ref{lateLxLKMzwin} and \ref{earlyLxLKMzwin} illustrate the dependence of the
specific X-ray luminosity of the late-type and early-type galaxies, respectively, on redshift (left panels) and optical luminosity (right panels).
If we fit the late-type galaxies with the double power-law model
(Eqn.~\ref{double_power_law}), we find that, in units of $10^{40}$ ergs~s$^{-1}$, 
\beqa 
       L^{\rm LATE}_{\rm x40, hard} = (5.9\pm1.6)\left( \frac{L_K}{L_{K*}} \right)^{1.2\pm0.4} (1+z)^{3.6\pm1.2}, \nonumber \\
       L^{\rm LATE}_{\rm x40, soft} = (2.5\pm0.6)\left( \frac{L_K}{L_{K*}} \right)^{1.1\pm0.2} (1+z)^{2.2\pm1.0}.
\label{eq:LxLKlate}
\eeqa
The reduced $\chi^2$ (per degree of freedom) for these fits are 0.78 and 0.81 for the hard and soft bands,
respectively.
If we fit the early-type galaxies, we find that 
\beqa 
       L^{\rm EARLY}_{\rm x40, hard} = (3.0\pm0.6)\left(  \frac{L_K}{L_{K*}} \right)^{1.0\pm0.3} (1+z)^{3.9\pm0.8}, \nonumber \\
       L^{\rm EARLY}_{\rm x40, soft} = (1.8\pm 0.4)\left( \frac{L_K}{L_{K*}} \right)^{1.1\pm0.2}(1+z)^{3.1\pm0.8}.
\label{eq:LxLKearly}
\eeqa
In this case, the reduced $\chi^2$ values are 0.86 and 1.14 for the hard and soft bands, respectively.
The reduced $\chi^2$ values for the best-fit curves given in the remainder of the paper are all similarly good,
i.e., $\approx 1\pm 0.2$.
In all cases, the upper limits set by non-detections (bins with S/N $ < 1$) are used in conjunction
with detections to establish the best-fit curves. In particular, if
any best-fit curves exceed these upper limits, the squared difference is added to the $\chi^2$ sum.

The results for the individual luminosity and redshift bins are summarized in Tables 2 and 3.
In spite of the evidence that different physical processes dominate 
the emission of the late-type and early-type galaxies, the two populations 
exhibit qualitatively similar behaviors. In particular,

\noindent\textbullet ~ the X-ray luminosity monotonically increases 
from low mass galaxies at low redshift to high mass galaxies at high redshift and  

\noindent\textbullet ~ the X-ray luminosity is significantly higher than 
what would be expected from LMXBs on the basis of the $K$-band luminosities, Eqn.~(\ref{eq:LxLK}),
assuming a hardness ratio of HR~$=0$ (e.g., Muno et al. 2004).

These similarities are not surprising, as we expect the X-ray emission of LMXBs and hot gas
to be nearly time-independent for $z \ltorder 0.5$,
while the HMXB and AGN emission should be evolving rapidly
and at approximately the same rate ($\propto (1+z)^{3\pm 1}$, e.g.,
David, Jones, \& Forman 1992; Shapley et al. 2001, Hogg 2001, Norman et al. 2004, and Schiminovich et al. 2005
and Ba01 and Ba05, respectively).
\subsection{Comparisons to Previous Work}
In general, we find excellent agreement between our results and those of earlier studies
of normal galaxy X-ray evolution, but our uncertainties are smaller.
There are three general categories of earlier results: surveys of nearby, individual galaxies, stacking analyses of
low to intermediate-redshift galaxies and stacking analyses of intermediate to high-redshift
galaxies in small-area, deep fields (CDF-N and CDF-S).  
Taking the late-type and early-type galaxies in turn, we first consider the
Shapley et al. (2001, S01) and O'Sullivan et al. (2001, OS01) surveys
of local spiral and early-type galaxies.
Second, we compare our
results to the Georgakakis et al.~(2003, G03) stacking analysis of 2dF galaxies at $z\simeq 0.1$
and to our own earlier stacking analysis of massive early-type galaxies 
at $0.3 \ltorder z \ltorder 0.9$ (B05).  
Finally, we consider the Hornschemeier et al. (2002, H02) and Lehmer et al. (2007) 
stacking analyses of normal galaxies in the CDF-N 
at $0.4 \ltorder z \ltorder 1.5$ and optically bright early-type galaxies in the extended CDF-S
out to $z \ltorder 0.7$, respectively.
  
Shapley et al. (2001, S01) studied the
correlation between the X-ray (0.2--4.0~keV) and $H$-band flux of
234 local spiral galaxies with
no sign of AGN activity.
Their results indicate a trend of
\beq
L^{\rm LATE}_{\rm x40,\rm S01} (z\simeq 0) \simeq (2\pm 1)\left(\frac{L_K}{L_{K*}} \right)^{1.2\pm0.1},
\label{eq:S01LxLK}
\eeq 
assuming H$-$K$=0.27$~mag
and a 50\% uncertainty in the normalization of the correlation. It is clear
that Eqn.~(\ref{eq:S01LxLK}) agrees very well with Eqn~(\ref{eq:LxLKlate}) at $z = 0$, and, as
shown in Fig.~\ref{lateLxLKMzwin}, 
the agreement extends to all redshifts if we allow for evolution (see Fig. 9).
Similar trends have been found in other local
surveys (e.g., David, Jones, \& Forman 1992). 

Georgakakis et al.~(2003, G03) considered 200 galaxies in the 2dF
redshift survey with $\langle L_B \rangle \simeq L_*$ at a mean
redshift $\langle z \rangle\simeq 0.1$.  They measured mean
soft band luminosities of $\log (L_{\rm x}/\rm{ergs~s}^{-1}) \simeq 40.2\pm 0.2$, $39.7 \pm 0.4$ and $40.5\pm 0.2$
for the combined, late-type and early-type
samples. They also estimated an upper limit of $\log(L_{\rm x}/\rm{ergs~s}^{-1}) \ltorder 40.5$ on
the hard band luminosities of the samples. These results are 
consistent with ours, particularly if we allow
for positive redshift evolution of the X-ray luminosity between $\langle z \rangle\simeq 0.1$ for G03 and
our points for galaxies of similar optical luminosities at $z\simeq 0.25$ and $0.45$.  

Hornschemeier et al. (2002, H02) conducted a stacking analysis of late-type galaxies
in the HDF-N/CDF-N, finding that 29,
$\simeq 0.7L_*$ galaxies in the redshift range $0.4-0.75$ have mean total ($0.5-8$~keV) 
and soft band ($0.5-2.0$~keV) luminosities of $L_{\rm x40}=2.9\pm0.4$ 
and $1.3\pm0.2$, respectively.  These too are consistent with the values we find
for $\simeq 0.7L_*$ galaxies at $0.2 \ltorder z \ltorder 0.3$ ($3.6\pm0.8$ 
and $1.6\pm0.3$ respectively).

\onecolumn
%
%FIG8: L_x vs. L_K at fixed z
%
\bfignoc 
\centerline{\epsfxsize=14.7cm \epsfbox{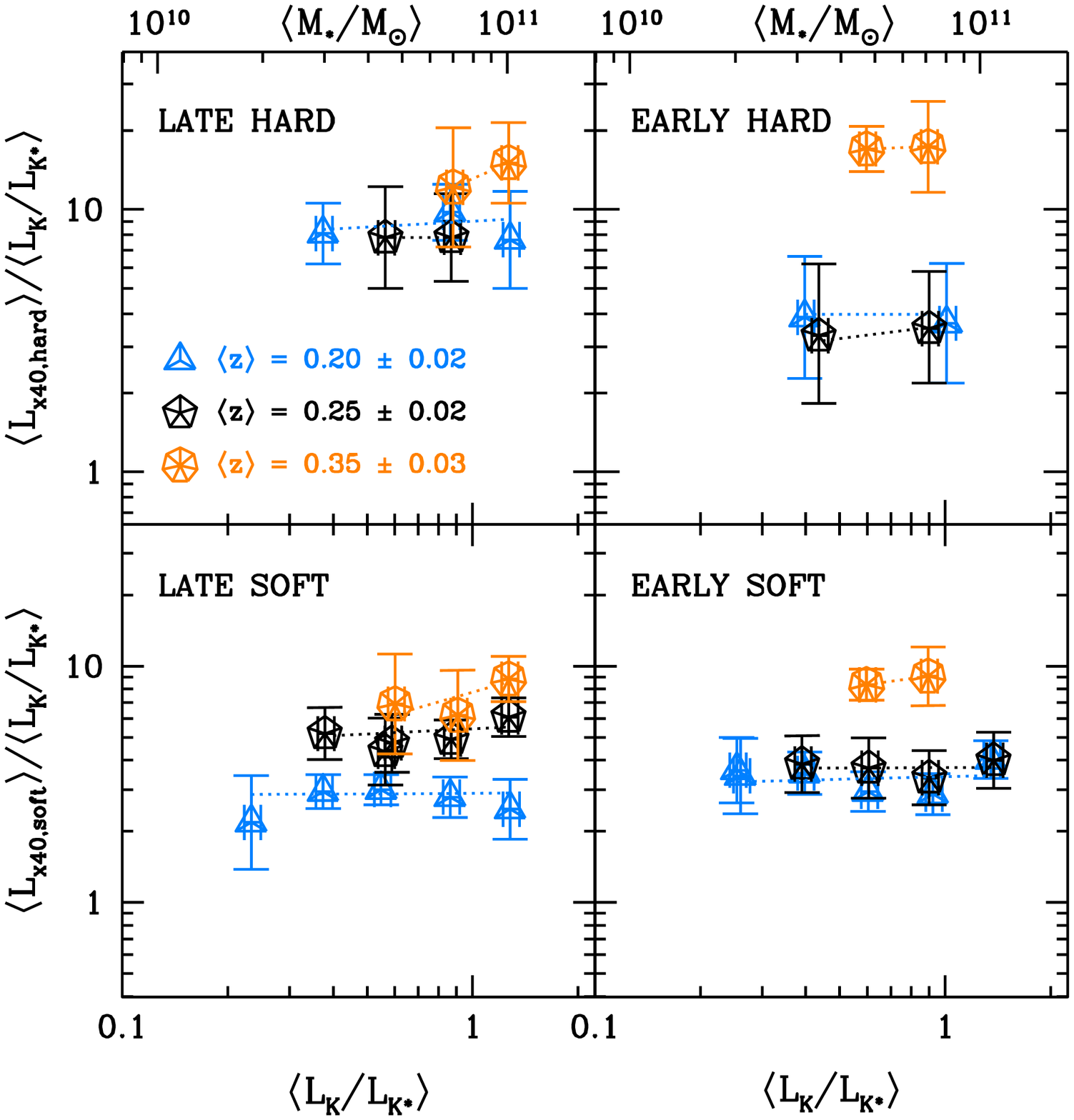}}
\caption{
The specific X-ray luminosity evolution of the early and late-type galaxies
in three narrow redshift slices: $\langle z \rangle = 0.20 \pm 0.02$ (three-pointed stars), $\langle z \rangle = 0.25 \pm 0.02$ (five-pointed stars),
and $\langle z \rangle = 0.35 \pm 0.03$ (seven-pointed stars).  In all redshift slices, the specific X-ray luminosities are
relatively flat in $L_K$ (roughly,
$L_{\rm x,\rm late} \propto L_K^{1.2 \pm 0.3} \propto L_{\rm x,\rm early}$, see \S4.3),
indicating an approximately linear correlation between $L_{\rm x}$ and $L_K$
as found previously (e.g., David, Jones, \& Forman 1992, Shapley et al. 2001).
}
\label{fixedzLxLK}
\efignoc

%
%FIG9: L_x vs. z and L_K for late-type sample     
%
\bfignoc
\centerline{\epsfxsize=14.7cm \epsfbox{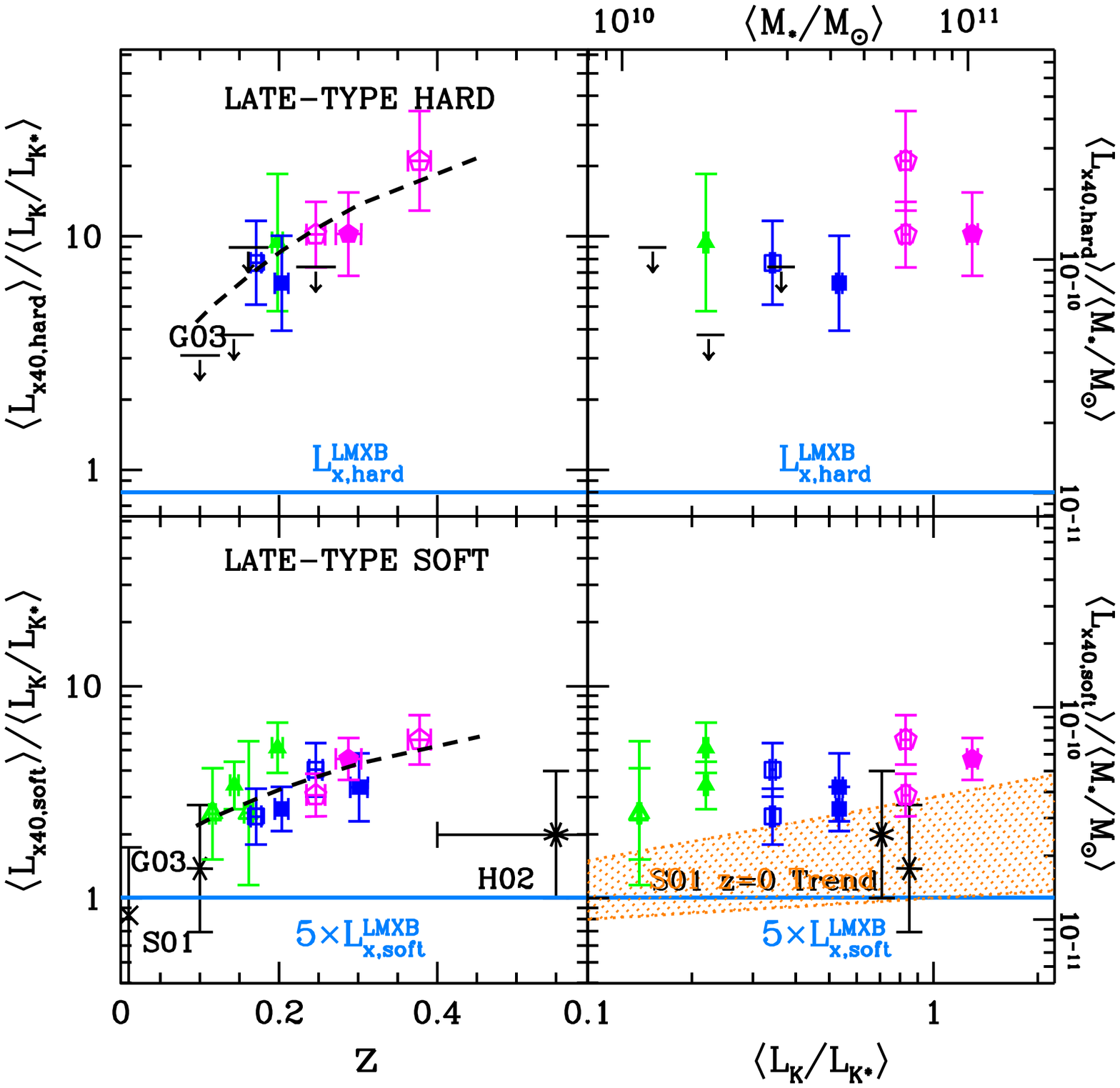}} %one col
\caption{ \bf{Left:} \rm
The mean hard and soft band specific X-ray luminosities of the late-type sample of galaxies 
as a function of redshift. 
The point type conventions are the same as those used in Figure \ref{randhspecLx}.
The best-fit curves for the hard and soft band emission are also shown.
\bf{Right:} \rm The mean hard and soft specific X-ray luminosities
as a function of $\langle {L}_{\rm K}\rangle$ for the late-type sample of galaxies. 
The instances in which there are two points at
each value of $L_K$ demonstrate the evolution from the low to the high redshift bin
in the corresponding absolute magnitude strip (Fig. 1). 
Top and right axes display the corresponding
behavior of $\langle L_{\rm x}\rangle/\langle {M}_{*}\rangle$ vs. $\langle M_{*}\rangle$ based 
on Eqn.~\ref{eq:Mstar}. Upper limits (1 $\sigma$) are provided in the hard band panels for
bins with negative net counts (see Table 2).
For comparison, we have plotted the
specific X-ray luminosities found by 
Shapley et al. (2001, S01, 4-pointed star at $z\simeq 0$),
Georgakakis et al. (2003, G03, 6-pointed star in the soft band panel;
upper limit in the hard band panel at $z\simeq 0.1$), and 
Hornschemeier et al. (2002, H02, 8-pointed star at $z\simeq 0.55 \pm 0.15$). 
The shaded region in the lower right panel shows the local S01 
$L_{\rm x,soft}(L_{K})$ trend, roughly
$\log(L_{\rm x40}) = (0.3\pm 0.2)+ (1.2\pm0.1) \log(L_K/L_{K*})$, which encompasses both
the scatter in and error bars on the S01 data points.
}
\vspace{2.0cm}
\label{lateLxLKMzwin}
\efignoc

\newpage

%
%FIG10: L_x vs. z and L_K for early-type sample
%
\bfignoc
\centerline{\epsfxsize=14.7cm \epsfbox{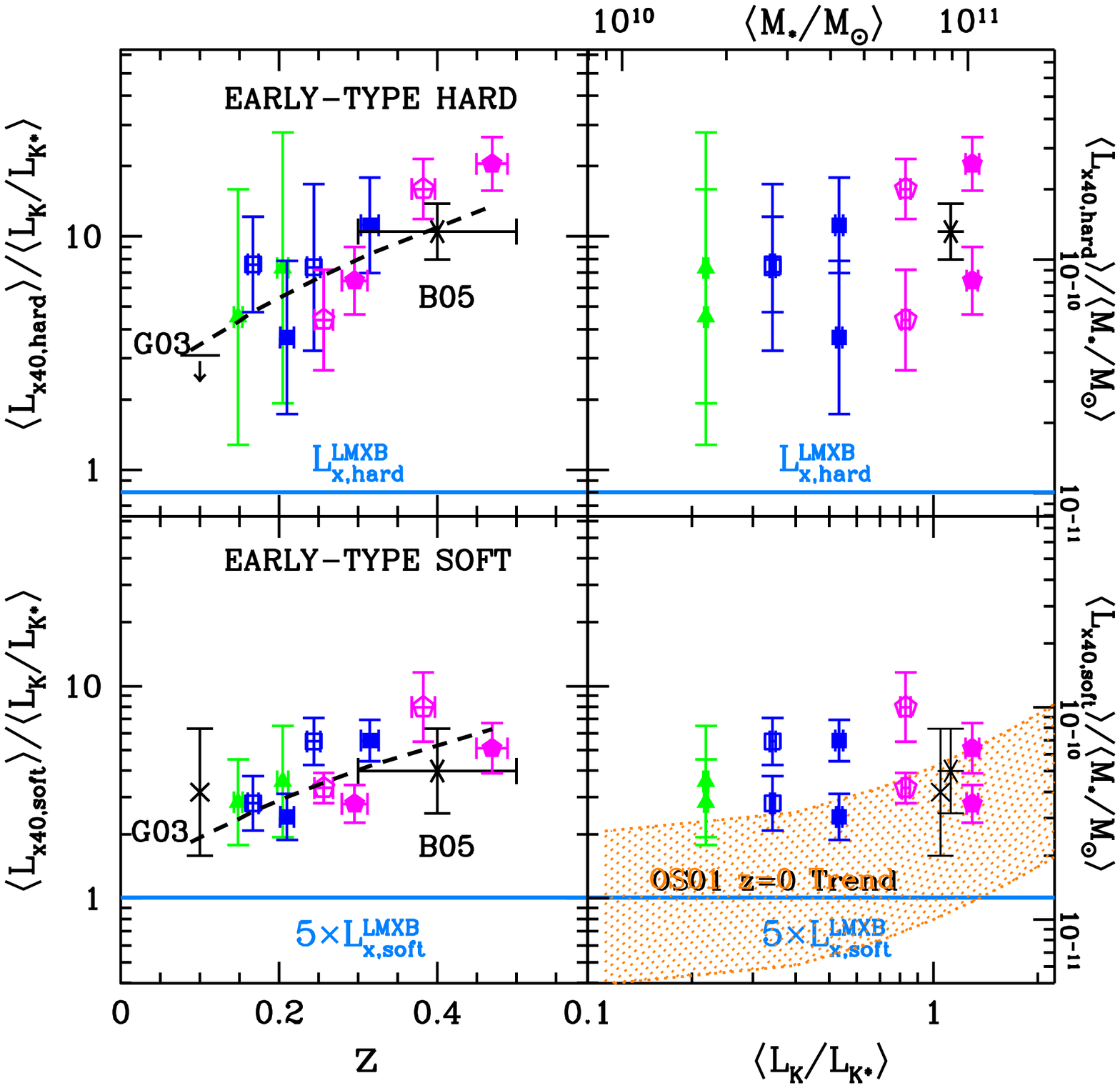}}  %one col
\caption{
Same as Figure \ref{lateLxLKMzwin} for the early-type sample of galaxies.
For comparison, we have plotted the specific X-ray luminosities of $z \simeq 0.1$ $L_*$ galaxies found by
Georgakakis et al. (2003, G03, 4-pointed star in the soft band panel; upper limit in the hard band panel) 
and Brand et al. (2005, B05, 6-pointed star in both panels at $z=0.4\pm0.1$).
The best-fit curve for soft and hard band emission from the AGES galaxies is also shown. 
The shaded region in the lower right panel shows the local O'Sullivan et al. (2001, OS01)
$L_{\rm x,soft}(L_{K})$ trend, roughly
$\log(L_{\rm x40}) = (0.3\pm 0.3)+ (1.7\pm0.2) \log(L_K/L_{K*})$, which encompasses both
the scatter in and error bars on the OS01 data points.
}
\label{earlyLxLKMzwin}
\vspace{5.0cm}
\efignoc
\twocolumn

\newpage

\onecolumn
\begin{deluxetable}{ccrrr}
\tablewidth{0pt}
\tablecaption{X-ray Counts From Stacked Normal Galaxies \& Excluded AGN}
\tablehead{
  Class &Source & \multicolumn{1}{c}{Number of} & \multicolumn{1}{c}{Net Hard}  & \multicolumn{1}{c}{Net Soft} \\
        &       & \multicolumn{1}{c}{Objects}   & \multicolumn{1}{c}{Counts}    & \multicolumn{1}{c}{counts} } 
\startdata
 Normal Galaxies   &Late-Type Galaxies & 3178  &240 &  327 \\  
                   &Early-Type Galaxies & 2968  &246 &  342 \\  
 Spectroscopic AGN &Broad Line AGN      &   27  &319 &  872 \\
                   &Narrow Line AGN     &   78  &496 &  404 \\
 X-ray Sources    &Broad Line AGN      &   24  &319 &  872 \\
                  &Narrow Line AGN     &   34  &468 &  367 \\
\enddata 
\label{tab:Xsources}
\end{deluxetable}

\def\h{\hphantom{0}}

\onecolumn
\begin{deluxetable}{crrrr}
\tablewidth{0pt}
\tablecaption{X-ray Luminosity Upper Limits for the Optically Faintest AGES Galaxies}
 \tablehead{
    $M_{\rm R,max}$
    &$M_{\rm R,min}$
    &$\langle z \rangle $
    &$N_{\rm galaxies}$
    &\multicolumn{1}{c}{$L_{\rm x40}$}\\
%%%%%%%%%%%%%%%%%%%%%%%%%%%%%%%%%%%%%%%%%%%%%%%% NEXT LINE IN COL HEADERs
   (mag)
    &\multicolumn{1}{c}{(mag)}
    &
    &
    &\multicolumn{1}{c}{(h/s)}
    }

\startdata
\multicolumn{5}{c}{Late-Type Galaxies} \\

$ -17.5$ &$ -18.5$ &$  0.0693$ &$  126$ &$  < 0.6$/$\h < 0.2$ \\
$ -17.5$ &$ -18.5$ &$  0.0951$ &$  126$ &$  < 1.9$/$\h < 1.0$ \\
$ -18.0$ &$ -19.0$ &$  0.0969$ &$  228$ &$  < 1.2$/$\h < 0.7$ \\
$ -18.0$ &$ -19.0$ &$  0.1286$ &$  228$ &$  < 1.2$/$\h < 0.4$ \\

\multicolumn{5}{c}{Early-Type Galaxies} \\

$ -17.5$ &$ -18.5$ &$  0.0756$ &$\h  13$ &$  < 1.8$/$\h < 0.4$ \\
$ -17.5$ &$ -18.5$ &$  0.0882$ &$\h  13$ &$  < 2.2$/$\h < 0.9$ \\
$ -18.0$ &$ -19.0$ &$  0.1051$ &$\h  59$ &$  < 2.6$/$\h < 0.6$ \\
$ -18.0$ &$ -19.0$ &$  0.1303$ &$\h  56$ &$  < 3.4$/$\h < 0.9$ \\
$ -18.5$ &$ -19.5$ &$  0.1166$ &$  119$ &$  < 2.4$/$\h < 0.4$ \\
$ -18.5$ &$ -19.5$ &$  0.1586$ &$  119$ &$  < 3.5$/$\h < 1.5$ \\

\enddata
\label{tab:Faint_Galaxies}
\tablecomments{
The range of absolute $R$-band magnitudes of the galaxies in a bin is $M_{\rm R, max} > M_R > M_{\rm R, min}$
and the mean redshift is $\langle z \rangle$.  $N_{\rm galaxies}$ is the effective
number of galaxies in the bin, including the correction for sparse sampling.  The 1 $\sigma$ upper limits
on the hard (h; 2--7~keV) and soft (s; 0.5--2~keV) band X-ray luminosities, $L_{\rm x40}$,
of the optically faintest AGES galaxies are given
in units of $10^{40}$~ergs~s$^{-1}$ in the final column.}
\end{deluxetable}

\begin{deluxetable}{crrrrrr}
\tablewidth{0pt}
\tablecaption{Summary of Results}
 \tablehead{ 
    $\langle M_R \rangle /\langle M_K \rangle$ 
    &$\langle z \rangle $ 
    &$N_{\rm galaxies}$ 
    &\multicolumn{1}{c}{Net Counts} 
    &\multicolumn{1}{c}{S/N} 
    &\multicolumn{1}{c}{$L_{\rm x40}$}
    &\multicolumn{1}{c}{$\langle L_{\rm x40}\rangle/\langle L_K/L_{K*}\rangle$}  \\
%%%%%%%%%%%%%%%%%%%%%%%%%%%%%%%%%%%%%%%%%%%%%%%% NEXT LINE IN COL HEADERs
   (mag)
    &
    &
    &\multicolumn{1}{c}{(h/s/t)} 
    &\multicolumn{1}{c}{(h/s/t)} 
    &\multicolumn{1}{c}{(h/s/t)} 
    &\multicolumn{1}{c}{(h/s/t)} 
    }
\startdata
\multicolumn{7}{c}{Late-Type Galaxies} \\

$ -18.98$/$   -21.31$ &$   0.1159$ &$ 375$ &$ \h 17$/$\h   34$/$\h   51$ &$   0.8$/$    2.0$/$    1.8$ &$ \h  0.7$/$\h    0.4$/$\h    1.1$ &$ \h  4.3$/$\h    2.5$/$\h    6.8  $\\
$ -19.02$/$   -21.33$ &$   0.1614$ &$ 376$ &$\h \h  -1$/$\h   17$/$\h   16$ & ---/$    1.3$/$    0.7$ &$ \h <1.4$/$\h    0.4$/$\h    0.3$ &$ \h <9.2$/$\h    2.5$/$\h    1.8  $\\
$ -19.42$/$   -21.75$ &$   0.1434$ &$ 580$ &$\h \h  -4$/$\h   71$/$\h   67$ & ---/$    3.9$/$    2.3$ &$ \h <0.8$/$\h    0.7$/$\h    0.6$ &$ \h <3.9$/$\h    3.4$/$\h    2.8  $\\
$ -19.47$/$   -21.80$ &$   0.1979$ &$ 576$ &$ \h 27$/$\h   52$/$\h   79$ &$   1.5$/$    3.6$/$    3.4$ &$ \h  2.1$/$\h    1.1$/$\h    3.2$ &$ \h  9.4$/$\h    5.1$/$   14.5  $\\
$ -19.93$/$   -22.25$ &$   0.1709$ &$ 667$ &$ \h 55$/$\h   61$/$  116$ &$   2.4$/$    3.3$/$    4.0$ &$ \h  2.6$/$\h    0.8$/$\h    3.5$ &$ \h  7.7$/$\h    2.4$/$   10.1  $\\
$ -19.94$/$   -22.26$ &$   0.2464$ &$ 667$ &$\h \h -23$/$\h   45$/$\h   22$ & ---/$    3.4$/$    1.0$ &$ <2.6$/$\h    1.4$/$ <3.0$ &$ <7.5$/$\h 4.0$/$   <9.0  $\\
$ -20.36$/$   -22.68$ &$   0.2032$ &$ 555$ &$ \h 39$/$\h   59$/$\h   98$ &$   2.1$/$    4.2$/$    4.2$ &$ \h  3.4$/$\h    1.4$/$\h    4.8$ &$ \h  6.3$/$\h    2.6$/$\h    8.9  $\\
$ -20.44$/$   -22.75$ &$   0.3011$ &$ 556$ &$\h \h 3$/$\h   30$/$\h   33$ &$   0.2$/$    2.7$/$    1.9$ &$ \h  0.6$/$\h    1.8$/$\h    2.4$ &$ \h  1.1$/$\h    3.3$/$\h    4.4  $\\
$ -20.85$/$   -23.16$ &$   0.2465$ &$ 403$ &$ \h 46$/$\h   50$/$\h   96$ &$   3.1$/$    4.3$/$    5.1$ &$ \h  8.5$/$\h    2.5$/$   11.0$ &$  10.2$/$\h    3.1$/$   13.2  $\\
$ -20.90$/$   -23.20$ &$   0.3774$ &$ 403$ &$ \h 36$/$\h   34$/$\h   70$ &$   3.6$/$    3.7$/$    5.1$ &$  17.5$/$\h    4.6$/$   22.2$ &$  21.1$/$\h    5.6$/$   26.6  $\\
$ -21.27$/$   -23.57$ &$   0.2878$ &$ 195$ &$ \h 25$/$\h   39$/$\h   64$ &$   2.4$/$    4.4$/$    4.7$ &$  13.2$/$\h    5.9$/$   19.1$ &$  10.2$/$\h    4.5$/$   14.8  $\\

\multicolumn{7}{c}{Early-Type Galaxies} \\

$ -19.55$/$   -21.88$ &$   0.1482$ &$ 328$ &$ \h    14$/$\h 31$/$\h 45$ &$   0.8$/$    2.2$/$    2.0$ &$ \h1.0$/$\h  0.6$/$\h  1.6$ &$ \h4.5$/$\h  2.8$/$\h  7.3  $\\
$ -19.51$/$   -21.84$ &$   0.2047$ &$ 327$ &$ \h    11$/$\h 19$/$\h 30$ &$   0.7$/$    1.7$/$    1.6$ &$ \h1.6$/$\h  0.8$/$\h  2.4$ &$ \h7.3$/$\h  3.6$/$   10.9  $\\
$ -19.99$/$   -22.31$ &$   0.1672$ &$ 560$ &$ \h    47$/$\h 62$/$  109$ &$   2.1$/$    3.4$/$    3.8$ &$ \h2.6$/$\h  1.0$/$\h  3.6$ &$ \h7.6$/$\h  2.8$/$   10.4  $\\
$ -20.00$/$   -22.32$ &$   0.2438$ &$ 561$ &$ \h    20$/$\h 53$/$\h 73$ &$   1.2$/$    3.9$/$    3.4$ &$ \h2.5$/$\h  1.9$/$\h  4.4$ &$ \h7.4$/$\h  5.5$/$   12.8  $\\
$ -20.45$/$   -22.77$ &$   0.2102$ &$ 804$ &$ \h    31$/$\h 72$/$  103$ &$   1.3$/$    4.0$/$    3.5$ &$ \h2.0$/$\h  1.3$/$\h  3.3$ &$ \h3.7$/$\h  2.4$/$\h  6.1  $\\
$ -20.45$/$   -22.76$ &$   0.3146$ &$ 796$ &$ \h    37$/$\h 65$/$  102$ &$   2.1$/$    4.5$/$    4.5$ &$ \h6.0$/$\h  3.0$/$\h  8.9$ &$  11.2$/$\h  5.5$/$   16.7  $\\
$ -20.90$/$   -23.21$ &$   0.2564$ &$ 844$ &$ \h    38$/$  103$/$  141$ &$   2.0$/$    6.1$/$    5.6$ &$ \h3.6$/$\h  2.8$/$\h  6.4$ &$ \h4.4$/$\h  3.3$/$\h  7.7  $\\
$ -20.90$/$   -23.20$ &$   0.3823$ &$ 844$ &$ \h    55$/$\h 98$/$  153$ &$   3.4$/$    6.7$/$    7.0$ &$  13.3$/$\h  6.6$/$   19.9$ &$  15.9$/$\h  8.0$/$   23.9  $\\
$ -21.33$/$   -23.63$ &$   0.2954$ &$ 594$ &$ \h    45$/$\h 69$/$  114$ &$   3.0$/$    4.9$/$    5.5$ &$ \h8.4$/$\h  3.6$/$   12.0$ &$ \h6.5$/$\h  2.8$/$\h  9.2  $\\
$ -21.38$/$   -23.67$ &$   0.4691$ &$ 593$ &$ \h    48$/$\h 42$/$\h 90$ &$   3.8$/$    3.7$/$    5.3$ &$  26.5$/$\h  6.6$/$   33.1$ &$  20.4$/$\h  5.1$/$   25.5  $\\
 
\enddata
\label{tab:results}
\tablecomments{
The mean $R$-band and $K$-band
absolute magnitudes of the galaxies in a bin are $\langle M_R \rangle$ and
$\langle M_K \rangle$ and the mean redshift is $\langle z \rangle$.  $N_{\rm galaxies}$ is the effective
number of galaxies in the bin, including the correction for sparse sampling.  The Net Counts column gives
the number of photons observed above background in the hard (h; 2--7~keV), soft (s; 0.5--2~keV) and
total (t; 0.5--7~keV) bands;
the corresponding signal-to-noise ratio (S/N) of the measurements appears in
the adjacent column. (Our detection criterion is S/N$~>1$).
$L_{\rm x40}$ is the mean X-ray luminosity
in the hard, soft and total bands in units of
$10^{40}$~ergs~s$^{-1}$, and the final column is the specific X-ray luminosity,
$\langle L_{\rm x40,(\rm h,s,t)} \rangle / \langle L_K/L_{K*} \rangle $, for each band.
In the cases of negative net counts, 1 $\sigma$ upper limits have been provided. Only bins that contain $\geq$ 100 galaxies
and have net stacked emission with S/N $ > 1$ in at least one X-ray band
have been included in the table.}
\end{deluxetable}

\rm

\newpage

\twocolumn

%
%FIG11   L_x/L_K for L_* GALAXIES
%
\bfignoc
\centerline{\epsfxsize=8.7cm \epsfbox{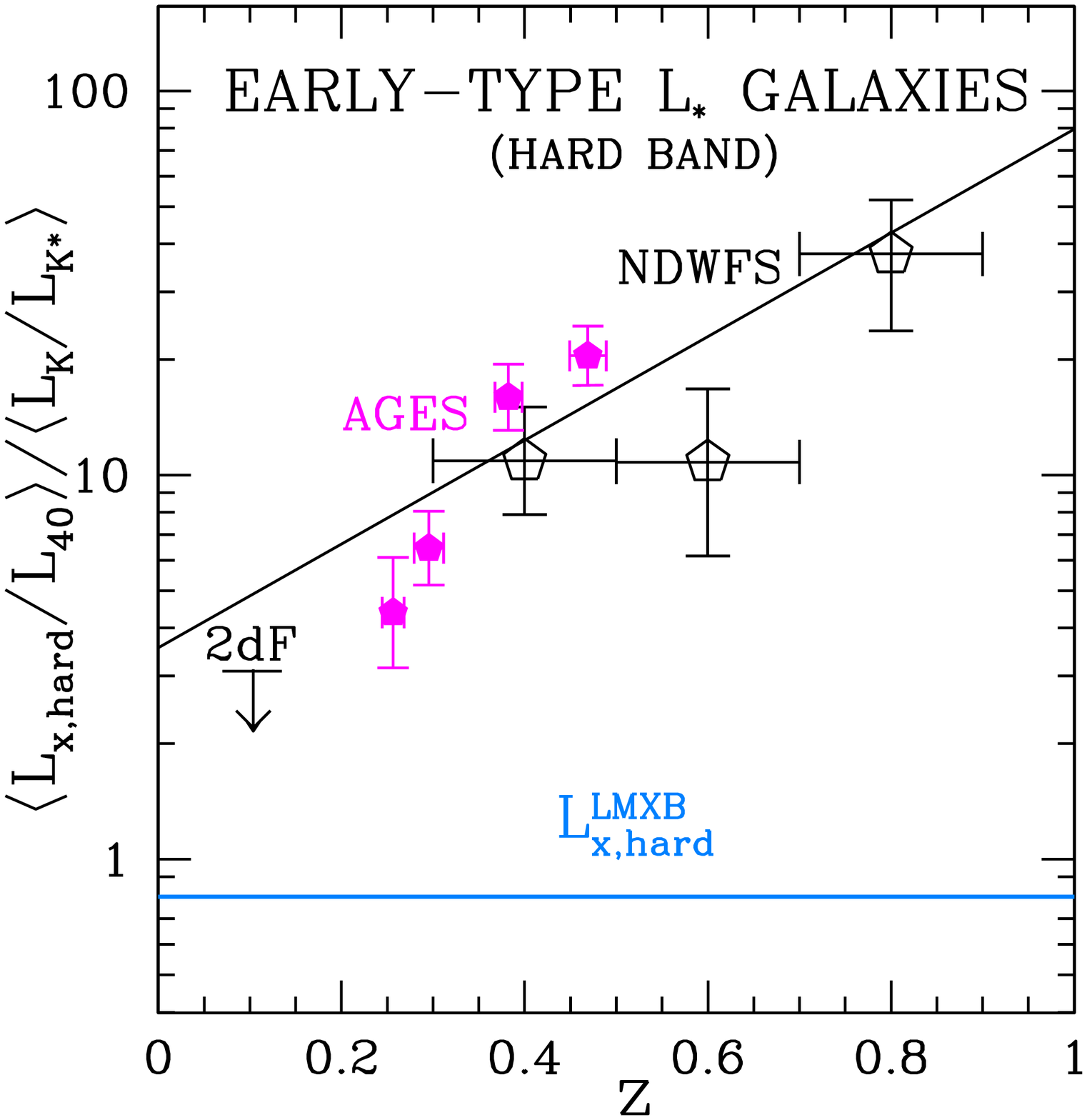}}
\vspace{0.5cm}
\centerline{\epsfxsize=8.7cm \epsfbox{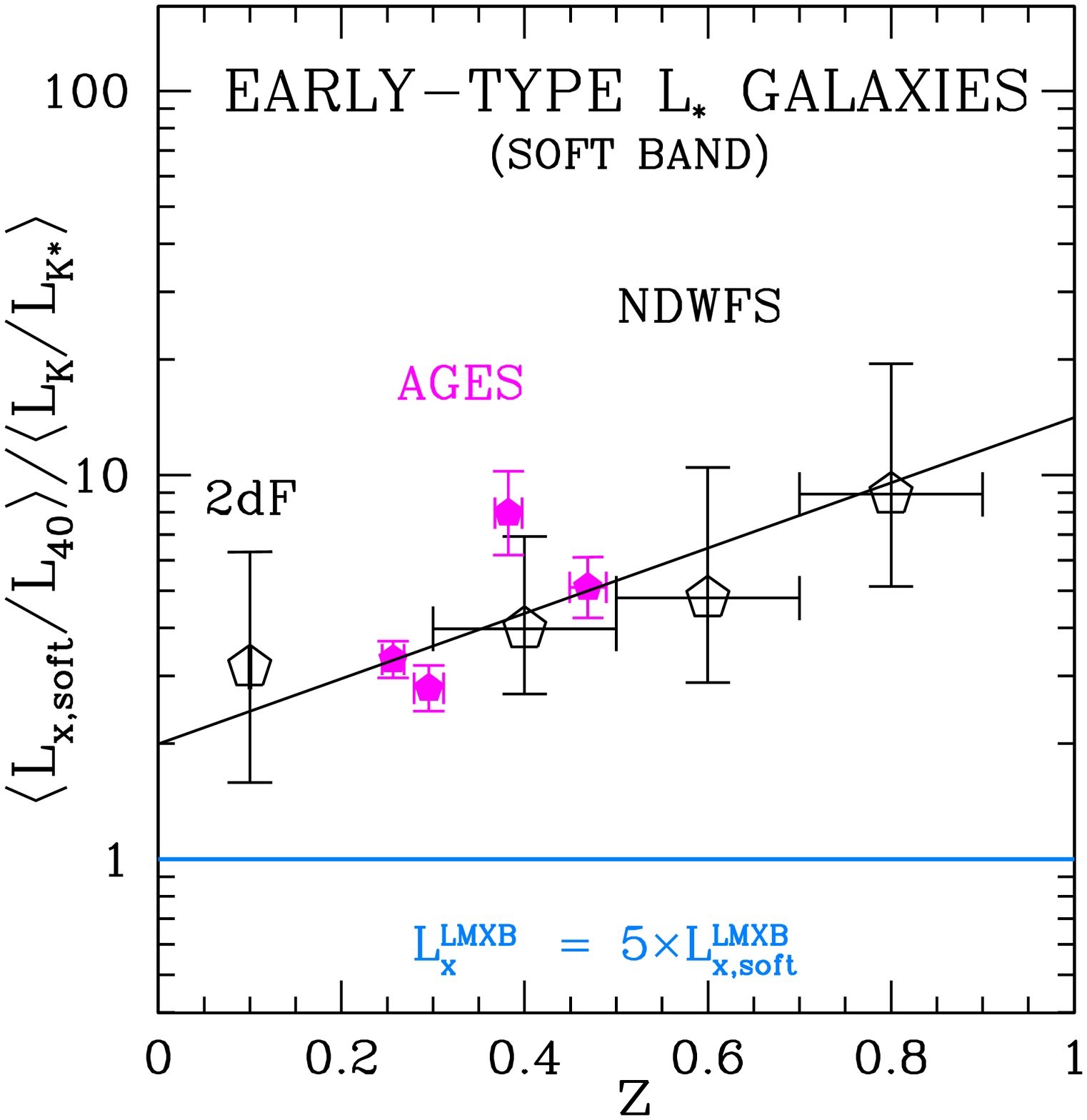}}
\caption{The mean hard (top) and soft (bottom) specific X-ray luminosities of $\simeq L_*$ early-type galaxies
from G03 (2dF, $z=0.1$), this work (AGES: filled points), and B05 (NDWFS: open 
points at $z \gtorder 0.4$).  
The solid lines are global fits to the combined samples (see \S4.4).
}
\label{LxLKstarearly}
\efignoc

%
%FIG12: SSFR
%
\bfignoc
\centerline{\epsfxsize=9.5cm \epsfbox{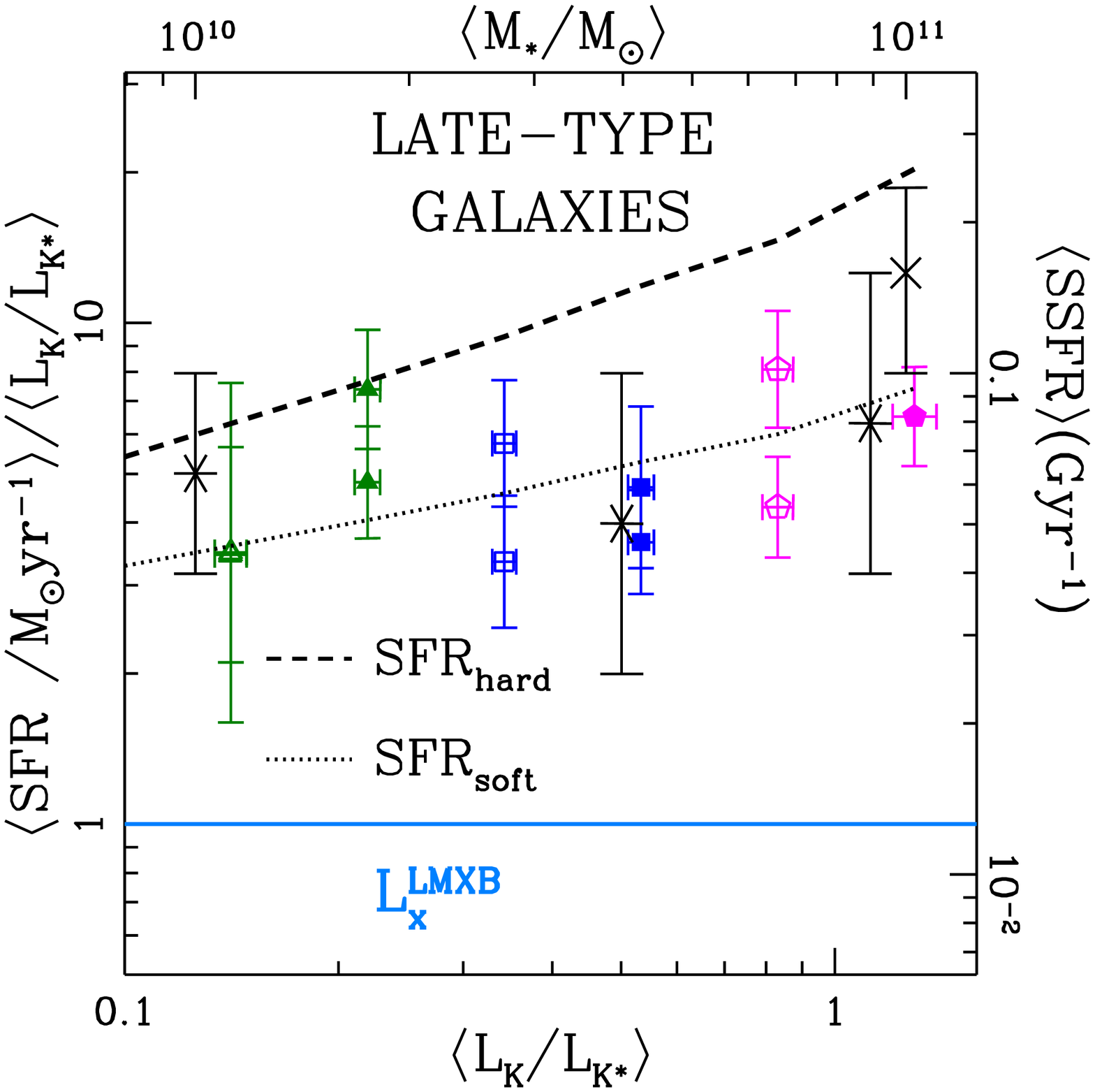}}
\caption{The Specific Star Formation Rate (SSFR = $\langle \rm SFR\rangle/\langle M_{*}\rangle$) 
as a function of $\langle L_K\rangle$ and $\langle M_{*}\rangle$ based on the
hard (dashed) and soft (dotted) X-ray luminosities of the AGES late-type galaxies and
Eqns. (\ref{eq:LxSFR}) - (\ref{eq:Mstar}).
For the soft band values, the point type conventions are the same as those used in Fig. 3. 
We compare our results to those of Bauer et al. (2005) (six-pointed stars),
and Bell et al. (2005) (four-pointed star).
}
\label{hspecSSFR}
\efignoc

%
%FIG13: SFR density
%
\bfignoc
\centerline{\epsfxsize=9.5cm \epsfbox{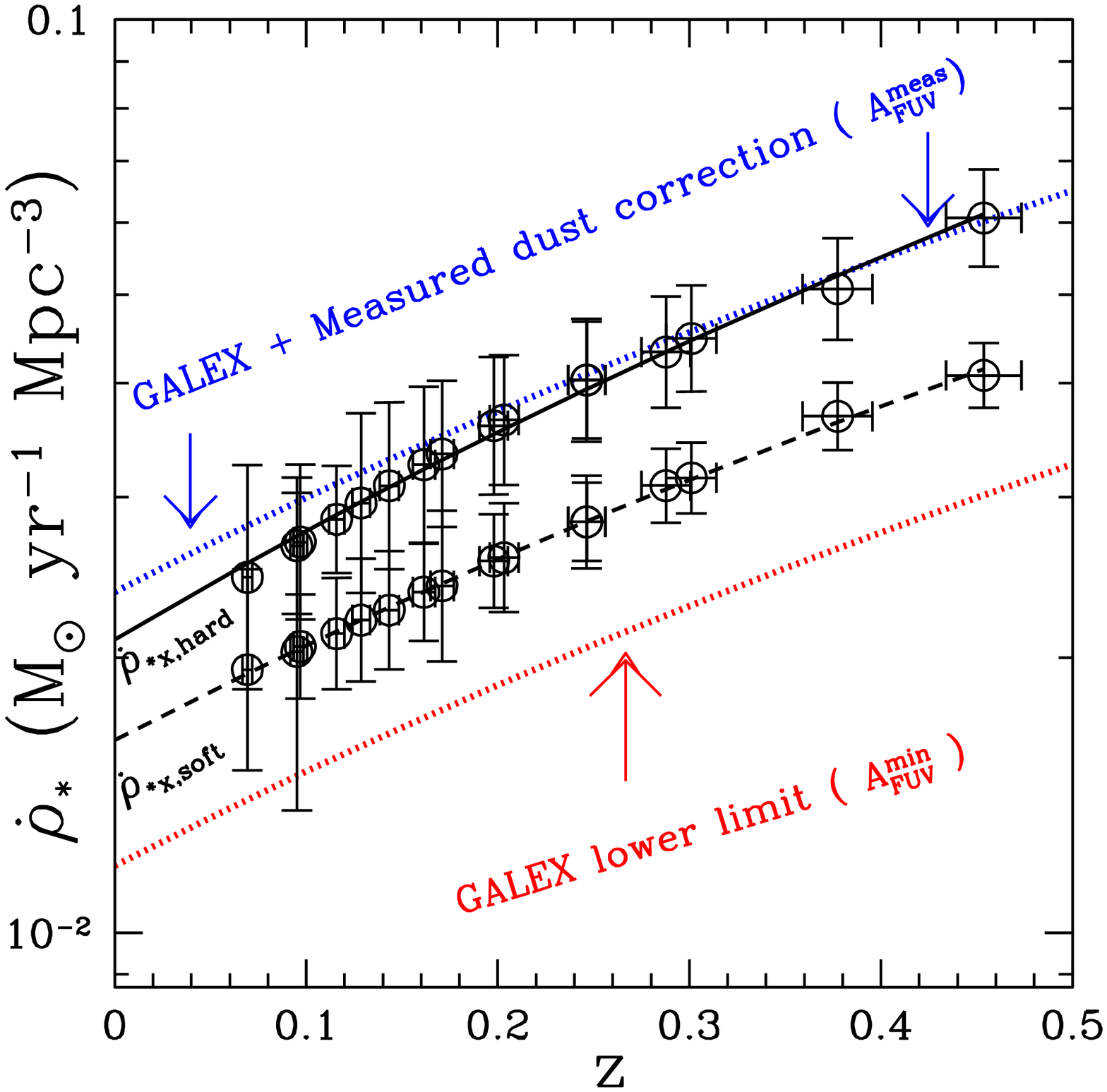}}
\caption{Estimates of the star formation rate density $\dot{\rho}_{*}$ based on the hard (solid)
and soft (dashed) X-ray luminosities of the AGES late-type galaxies as compared 
to the GALEX results (Schiminovich et al. 2004)
after either their minimum ($A^{\rm min}_{\rm FUV}$) or best-fit ($A^{\rm meas}_{\rm FUV}$)
corrections for dust extinction have been applied.}
\label{hspecSFRd}
\efignoc

%
%FIG14  \dot{M}_BH(z) VS. M*   
%
\bfignoc
\centerline{\epsfxsize=9.5cm \epsfbox{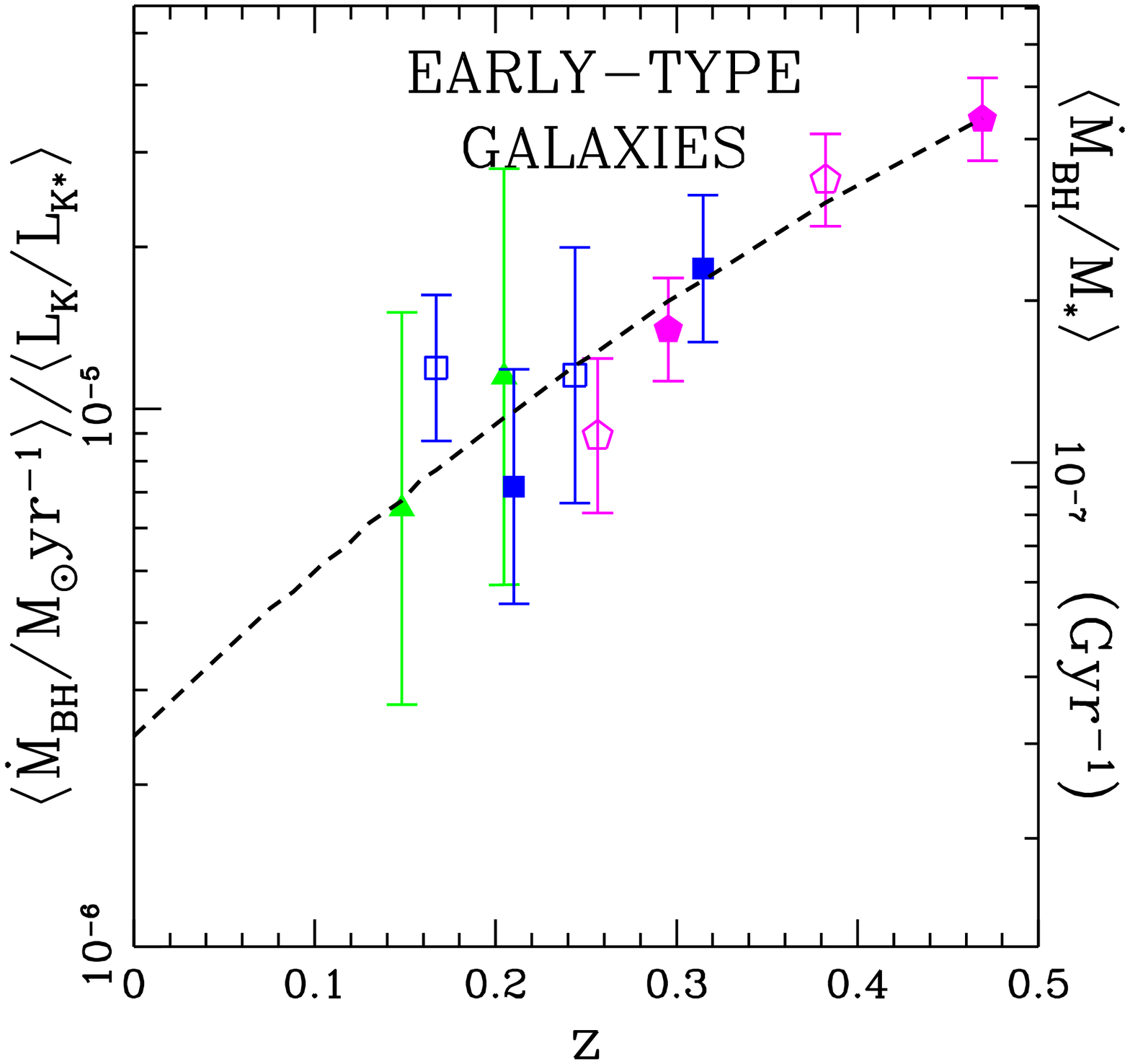}}
\caption{The specific supermassive black hole (SMBH) accretion rate 
as a function of redshift based on the hard band X-ray 
luminosity of AGES early-type galaxies (in excess of LMXB emission) and Eqn.~(\ref{eq:Mdot}). 
The dashed line is the curve of best fit.
The point type conventions are the same as those used in Fig. 3.
}
\label{hspecMdoth}
\efignoc
%
%FIG15: SMBH accretion rate density
%
\bfignoc
\centerline{\epsfxsize=9.5cm \epsfbox{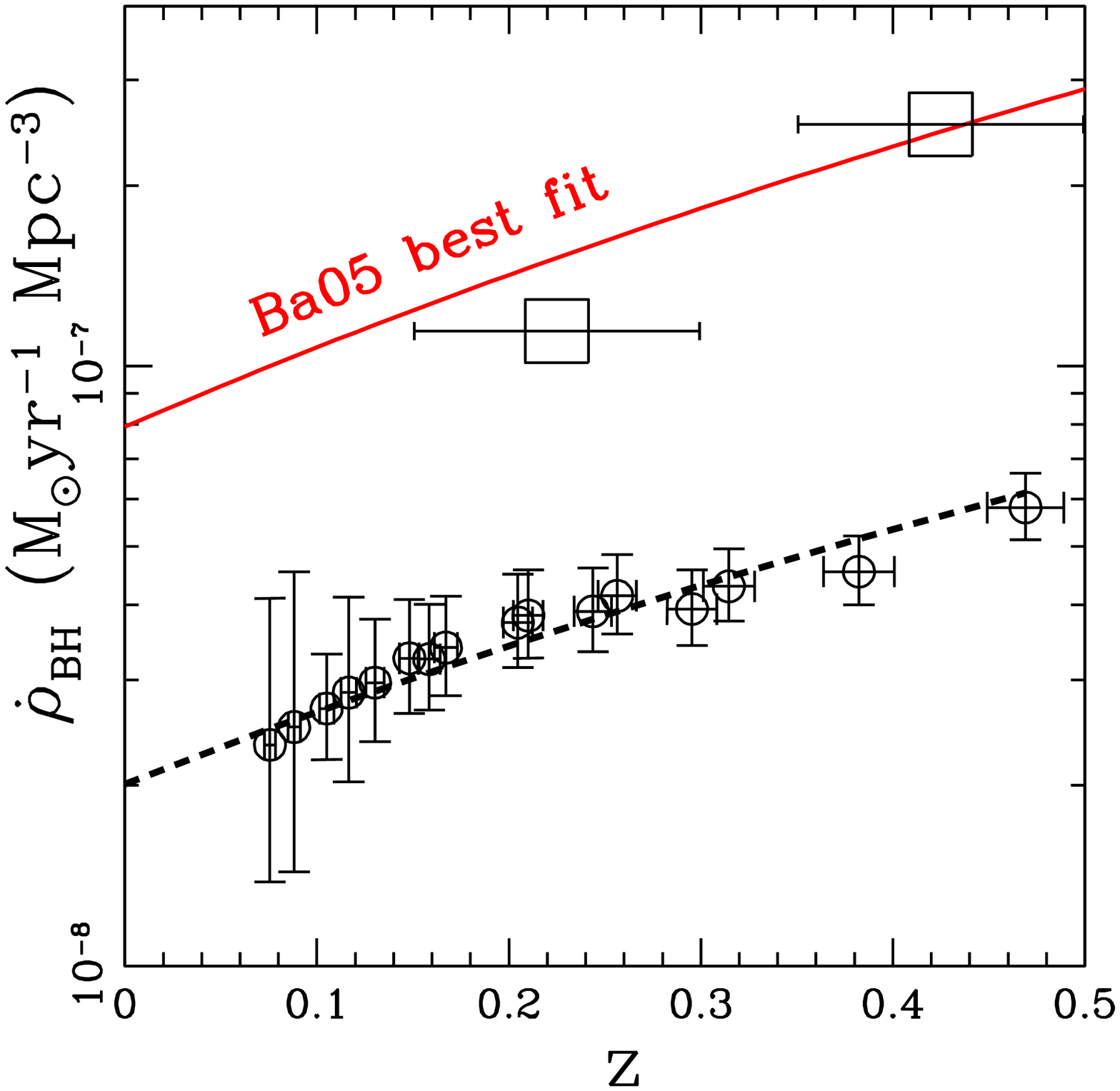}}
\caption{The comoving supermassive black hole (SMBH) accretion rate density as a function of 
redshift for the AGES early-type galaxies.  Because we are dealing with normal galaxies
rather than X-ray bright AGN, our results lie roughly a factor of four below the 
Ba05 data (open squares at $0.15 < z < 0.3$ and $0.35 < z < 0.5$) and the
(solid) Ba05 best-fit curve (which is based on additional data out to $z\simeq 1.2$). 
The slopes of the best-fit curves shown do agree to within errors, however.
Our value is $2.9 \pm 0.7$ (dashed line), while Ba05 finds $3.2 \pm 0.8$ (for $z < 1.2$). 
}
\label{hspecMBHd}
\efignoc

Turning to the early-type galaxies, we first checked the
relative contributions of hot gas and AGN emission by comparing our results
to the ROSAT~PSPC (0.1-2.5 keV) X-ray survey of
401 local early-type galaxies
discussed in O'Sullivan et al. (2001).
Because the hot gas emission from the early-type galaxies is primarily a
function of galaxy mass rather than epoch,
there should be little difference between the soft band flux of the local and AGES
galaxies (at $0.15 \ltorder \langle z \rangle \ltorder 0.45$),
provided that the soft band AGN contamination in the AGES sample is small.
Fig.~\ref{earlyLxLKMzwin} shows that the soft band luminosity
from the AGES early-type galaxies is, in fact, comparable to that of the local sample
even at high optical luminosities (which tend to be associated with
higher redshifts in the AGES sample). In contrast,
the total luminosity of the AGES early-type galaxies is consistently
higher than that of the local sample.  Fig.~\ref{earlyLxLKMzwin} also demonstrates that the
hard X-ray component of the early-type galaxies is
well in excess of the estimated LMXB emission and
undergoes rapid, positive redshift evolution.
These findings reinforce our conclusions in \S4.2, that the soft band flux of the AGES early-type
galaxies comes primarily from hot gas, while the more rapidly evolving
hard component comes primarily from AGN.

Our results for the massive ($\sim L_*$) early-type galaxies agree
with those of G03 and with our earlier work (B05), where we found that bright red ($M_R < -21.3$)
galaxies at $z\simeq 0.4$ had total X-ray luminosities of $L_{\rm x} \simeq 10^{41}\rm{ergs~s}^{-1}$. 
By incorporating the results of G03 and B05 with ours, we gain more leverage on the behavior of $L_*$ early-type
galaxies at lower ($z\simeq 0.1$) and higher ($0.5\ltorder z \ltorder 0.9$) redshifts
than what we can probe with AGES alone. 
Using this combined sample, we find that the hard and soft band luminosities of 
$L_*$ early-type galaxies follow the trends
\linebreak[4]
\beq
L^{\rm EARLY}_{\rm x40,\rm hard}(L_*) = (3.6\pm1.5)(1+z)^{4.5\pm1.7}
\label{eq:LxhLKstarearly}
\eeq
and
\beq
L^{\rm EARLY}_{\rm x40,\rm soft}(L_*) = (2.0\pm 0.8)(1+z)^{2.8\pm 1.1}
\label{eq:LxsLKstarearly}
\eeq
\linebreak[4]
out to $z\simeq 1$. 
These global fits, which are shown in Fig.~\ref{LxLKstarearly},
are consistent with those found for the
AGES $L_*$ early-type galaxies alone (Eqn.~\ref{eq:LxLKearly} at $L_K = L_{K*}$). 

Lehmer et al. (2007) conducted a stacking analysis of 222 optically bright
($\sim L_{B*}$) early-type
galaxies in the E-CDF-S with photometric redshifts ranging from 0 to 0.7.  The mean hard X-ray
luminosity trend inferred for AGN in the Lehmer et al. galaxy sample is in excellent agreement
with the hard band trend found in B05 (see Lehmer et al. 2007; Fig. 13) and in the present paper
(Fig.~\ref{LxLKstarearly}).

\section{Interpretation as Star Formation and Nuclear Accretion}

In this section we interpret the trends in X-ray luminosity in terms of star formation (HMXBs)
and nuclear accretion (AGN) after estimating and subtracting the additional contributions of LMXBs and hot gas.  
Under the assumption, discussed in the introduction, that Eqn.~\ref{eq:LxLK} holds out to $z\simeq 0.5$,
LMXBs represent a modest contribution to the emission from the late-type ($\ltorder 15\%$)
and early-type ($\ltorder 20\%$) galaxies.
The comparison of the (primarily) hot gas emission from local early-type galaxies
(O'Sullivan et al. 2001) shown in Fig.~\ref{earlyLxLKMzwin} suggests that, at all redshifts,
$\simeq 50\%$ of the soft band (and $\ltorder 20\%$ of the total) emission
of the AGES early-type galaxies is produced by hot gas.
Based on these arguments, we conclude that $\gtorder 85\%$ of the X-ray flux
of the late-type galaxies is produced by HMXBs and
$\gtorder 60\%$ of the total (and $\gtorder 80\%$ of the hard) X-ray flux of
the early-type galaxies is produced by low luminosity AGN.
In the analysis that follows, we use the hard and soft band luminosities of the
late-type galaxies to
estimate the evolution of star formation rates in normal galaxies and the hard band
luminosity of the early-type galaxies to estimate
the growth rates of supermassive black holes in normal galaxies.

Before we can implement these assumptions, we must also consider
the possibility that the flux from the high redshift late-type galaxies is being contaminated by AGN.
Since we cannot resolve their radial emission profiles as we did for the nearby subsamples (\S4.1)
and their luminosity evolution is fairly similar to that of the early-type galaxies, which we largely attribute to
AGN, we must find other ways to address this possibility.
Fortunately, there are three key properties of the late-type galaxy emission
that suggest the level of AGN contamination is small.
First, the star formation signatures in the late-type galaxy optical spectra are necessarily
correlated with a significant flux from HMXBs (Eqn. 1), suggesting that
only a small fraction of the late-type galaxy emission could come from AGN.
Second, the X-ray emission from late-type galaxies
is extended at low redshifts (Fig. 6), directly demonstrating the absence
of AGN among the nearby members of the sample.
Third, the Poisson and bootstrap error estimates are similar
at all redshifts, which they would not be if a small number of individual,
high-count sources (AGN) were hidden within the sample.
This disfavors small to moderate AGN contamination.
Significant AGN contamination is also disfavored by
our analysis and comparisons to previous work in \S5.1.

\subsection{Star Formation Rates in the Late-Type Galaxies}

Studies of star formation over the redshift range $0\ltorder z \ltorder 3$ have demonstrated
that the relationship bewteen X-ray luminosity and star formation rates is reasonably
universal (e.g., Brandt et al.  2001; Seibert et al. 2002; Nandra et al. 
2002; Cohen 2003; Grimm et al. 2003; Persic et al. 2004; Reddy \& Steidel 2004; 
Lehmer et al. 2005).  We use Eqn.~(\ref{eq:LxSFR}) to estimate the hard band SFR 
from $L^{\rm HMXB}_{\rm x, hard} = L^{\rm LATE}_{\rm x, hard}-L^{\rm LMXB}_{\rm x, hard}$, where
$L^{\rm LMXB}_{\rm x, hard}$ is based on Eqn. (2) and a typical LMXB spectrum with $HR=0$ (e.g., Muno et al. 2004)
and $L^{\rm LATE}_{\rm x, hard}$ is given in Eqn.~(\ref{eq:LxLKearly}).  The result is
\linebreak[4]
\beqa
   \hbox{SFR}_{\rm hard}= (6.4\pm1.8)\langle L_K/L_{K*} \rangle^{1.2\pm0.4}\nonumber \\
%\nonumber
\times (1+z)^{3.4\pm1.2} \hbox{M}_\odot \hbox{yr}^{-1}.
\label{eq:SFRhard}
\eeqa
\linebreak[4]
Using Eqn.~(\ref{eq:Mstar}) to estimate the stellar mass $M_*$ 
%\linebreak[4]
of the galaxies, we find that SFR$_{\rm hard}$ corresponds to a
%\linebreak[4]
specific star formation rate (SSFR = SFR/$M_*$) of
\linebreak[4]
\beqa
\hbox{SSFR}_{\rm hard} = (8.4\pm2.4)\times 10^{-2}(1+z)^{3.4\pm1.2} \nonumber \\
%\nonumber
\times \langle M_*/10^{11}M_{\odot} \rangle^{0.2\pm0.4}\hbox{Gyr}^{-1}.
\label{eq:SSFRhard}
\eeqa
\linebreak[4]

Based on the work of Ranalli et al. (2003, as modified by Gilfanov et al. 2004) we 
also use Eqn.~(\ref{eq:LxSFR}) to estimate the soft band SFR from $L^{\rm HMXB}_{\rm x, soft} =
L^{\rm LATE}_{\rm x, soft}-L^{\rm LMXB}_{\rm x, soft}$:
\linebreak[4]
\beqa
   \hbox{SFR}_{\rm soft}= (3.4\pm0.7)\langle L_K/L_{K*} \rangle^{1.1\pm0.2} \nonumber \\
%\nonumber
\times (1+z)^{2.4\pm1.0} \hbox{M}_\odot \hbox{yr}^{-1} 
\label{eq:SFRsoft}
\eeqa
\linebreak[4]
and a corresponding specific star formation rate of
\linebreak[4]
\beqa
   \hbox{SSFR}_{\rm soft} = (4.3\pm0.9)\times 10^{-2} (1+z)^{2.4\pm1.0} \nonumber \\
%\nonumber
\times \langle M_*/10^{11}M_{\odot} \rangle^{0.1\pm0.2}\hbox{Gyr}^{-1}.  
\label{eq:SSFRsoft} 
\eeqa
%\linebreak[4]

As shown in Fig.~\ref{hspecSSFR}, the results of earlier SSFR studies by Bauer et al.~(2005),
Bell et al.~(2005), and Feulner et al.~(2004) fall within the range encompassed by our
hard and soft band SSFR trends (Eqns.~\ref{eq:SSFRhard} and \ref{eq:SSFRsoft}),
but we have smaller uncertainties due to our larger sample size. 
Our results also agree with those of more recent 
analyses of comparably large galaxy samples (Noeske et al. 2007 and Zheng et al. 2007) in 
bins of similar stellar mass ($10.5 \ltorder \log(M_*/M_\odot) \ltorder 11$) and redshift 
($0.2 \ltorder z \ltorder 0.45$ and $z\simeq 0.3$, respectively).

We compute the star formation rate per comoving volume, $\dot{\rho}_{\rm *}$,
by integrating Eqns.~(\ref{eq:SFRhard}) and (\ref{eq:SFRsoft}) over the 
comoving density of the late-type galaxies (see \S2.2). These results are shown in Fig.~\ref{hspecSFRd}. 
We find
\beq
\dot{\rho}_{\rm *,hard}=(2.1\pm 0.4) \times 10^{-2}(1+z)^{2.9\pm 0.7}\frac{\rm M_\odot}{\rm yr~Mpc^{3}}
\label{eq:SFRdhard}
\eeq
and
\beq
\dot{\rho}_{\rm *,soft}=(1.6\pm 0.3) \times 10^{-2}(1+z)^{2.5\pm 0.5}\frac{\rm M_\odot}{\rm yr~Mpc^{3}},
\label{eq:SFRdsoft}
\eeq 
\linebreak[4]
for the hard and soft bands respectively.  As illustrated in Fig.~\ref{hspecSFRd}, our results
are broadly consistent with the range of $\dot{\rho}_{\rm *}$ values inferred from 
the standard extinction-corrected ultraviolet luminosity measurements of GALEX (Schiminovich et al. 2004).

It is important to note that significantly larger extinction corrections have already been ruled out 
by neutrino detectors. Values of $\dot{\rho}_*$ more than 50\% larger than $\dot{\rho}_{\rm *,hard}$ 
would be in serious conflict with the Super-Kamiokande upper limit (Malek et al. 2003) on the
Diffuse Supernova Neutrino Background (DSNB) from $M > 8 M_\odot$
stars (Strigari et al. 2005, Hopkins and Beacom 2006).
 
Our soft band estimates of $\dot{\rho}_*$ sit closer to the GALEX
minimum extinction correction curve. Nevertheless, the consistency of
our hard and soft band results with each other
and with the range of GALEX results indicates that
we are in general agreement with earlier studies 
(e.g., based on H$\alpha$ measurements: Tresse \& Maddox 1998, 
Tresse et al. 2002, P\'erez-Gonz\'alez et al. 2003, and Brinchmann et al. 2004
and the meta-analysis of Hogg 2001) as well as more recent comprehensive studies (e.g., Hopkins \& Beacom 2006). 

The level of agreement between our hard and soft band results also provides a check on the contamination
of our sample by AGN.  We first note
that the soft band SSFR and $\dot{\rho}_{\rm *}$ values (Eqns.~\ref{eq:SSFRsoft} and \ref{eq:SFRdsoft})
are systematically lower than the corresponding hard band values (Eqns.~\ref{eq:SSFRhard} and \ref{eq:SFRdhard}),
but this is not uncommon.
Many previous studies have also generally inferred lower SFR and $\dot{\rho}_*$ values based
on soft band flux (e.g., G03 and Norman
et al. 2004), suggesting that a low-normalization conversion between the soft band X-ray
flux and the star formation rate (e.g., Hornschemeier et al. 2005),
rather than contamination in the hard band, is to blame for this discrepancy.
Furthermore, if sources harder than X-ray binaries, like Type II AGN (see Fig. \ref{HR}),
were contributing excess hard band emission to the late-type galaxy flux, $\dot{\rho}_{\rm *,hard}$
(Eqn.~\ref{eq:SFRdhard}) would lie above rather than slightly below the
GALEX best-fit curve (Fig. \ref{hspecSFRd}). 

\subsection{Accretion Rates in the Early-Type Galaxies}

For the early-type galaxies, we use  Eqn.~(\ref{eq:LxLK}) to estimate and subtract
the contamination from LMXBs and use
the remaining hard X-ray luminosity to determine the average supermassive
black hole growth rate via Eqn.~(\ref{eq:Mdot}). 
Assuming a constant accretion efficiency of $\epsilon = 0.1$, we find that\\
\beqa 
\dot{M}_{\rm BH} = (5.8 \pm 1.1) \times 10^{-6} \langle L_{K}/L_{K*} \rangle^{1.1\pm0.4}\\
\nonumber
\times (1+z)^{3.3\pm 0.7} \rm M_\odot \rm yr^{-1}.
\label{eq:MBHdot}
\eeqa
\linebreak[4]
This growth rate (see Fig.~\ref{hspecMdoth}) is roughly an order of magnitude 
below the estimates of Ba01 ($\dot{M}^{\rm B01}_{\rm BH} \gtorder 10^{-4} M_\odot \rm yr^{-1}$),
because we are considering very different source populations.
The Ba01 estimates are based on directly detected X-ray sources with
luminosities exceeding $10^{42}$~ergs~s$^{-1}$, while we have eliminated such bright sources and
consider only ``normal'' galaxies with mean luminosities of order $10^{41}$~ergs~s$^{-1}$. 
The rapid redshift evolution of the accretion
rate in normal galaxies is consistent, however,
with the redshift evolution of bright AGN found by Ba01 and Ba05.

We estimate the accretion rate density $\dot{\rho}_{\rm BH}$ by integrating the accretion rate
per unit (stellar) luminosity over the early-type galaxy luminosity function (see \S2.2). 
\linebreak[4]
We find that\\
\beq
\dot{\rho}_{\rm BH}=(2.0\pm 0.4)\times 10^{-8}(1+z)^{2.9\pm 0.7}\rm M_\odot \rm yr^{-1} \rm Mpc^{-3}.
\label{eq:rhoMBH}
\eeq
\linebreak[4]
As shown in Fig.~\ref{hspecMBHd}, the evolution rate we determine is consistent with the rates found by Ba01
($\dot{\rho}_{\rm BH} \propto (1+z)^{3\pm 1}$) and Ba05 ($\dot{\rho}_{\rm BH} \propto (1+z)^{3.2\pm 0.8}$)
but we have a systematically lower normalization for $\dot{\rho}_{\rm BH}$ because of the different
source populations we have examined.  
This suggests that even at $z \ltorder 0.5$ the
aggregate growth rate of black holes in normal galaxies
is still a factor of a few less than that occurring in the much smaller number of
bright, low-redshift AGN.

\section{SUMMARY AND CONCLUSIONS}

After eliminating AGN flagged by AGES spectroscopy and X-ray sources brighter
than our luminosity limit (Eqn. 8), we used a stacking technique to determine the average 
X-ray properties of a magnitude-limited sample of 6146
($R<20$~mag, $z<0.6$) apparently normal galaxies.  
The results substantially improve our knowledge of normal galaxy X-ray evolution
over the redshift range $0.1 \ltorder z \ltorder 0.5$, which has been difficult to probe based on either
local studies of individual objects or small-area deep fields. 

We spectroscopically divided
the galaxies into late-type and early-type subsamples. 
The spectroscopic signatures of star formation,
spatially extended radial emission profiles, luminosity evolution, and low LMXB emission ($\ltorder 15 \%$)
of the late-type galaxies suggest that their X-ray emission
is dominated by HMXBs and therefore traces the star formation rate.

Conversely, because the early-type galaxies lack spectroscopic evidence for star 
formation, yet have rapidly evolving hard X-ray luminosities that are 
well in excess of that expected from LMXBs ($\ltorder 20 \%$), we concluded that their emission is
increasingly dominated by AGN at higher redshifts.

When we use a double power law ($L_{\rm x} \propto L_K^\alpha (1+z)^\beta$, Eqn.~\ref{double_power_law})
to fit the trends in the X-ray emission, we see that the X-ray luminosity increases 
monotonically from low mass galaxies at low redshift to high mass galaxies at high redshift
for both galaxy types. The redshift evolution of the late-type galaxies, 
$L_{\rm x} \propto (1+z)^{3\pm1}$, is in good agreement with previous 
estimates (e.g., H02, Ptak et al. 2001) and with theoretical expectations for 
normal, star-forming galaxies (Ghosh \& White 2001). The optical luminosity dependence of the late-type 
galaxy emission, roughly $L_{\rm x} \propto L_K^{1.2 \pm 0.3}$, also matches previous results
(e.g., S01; David, Jones, \& Forman 1992).
In addition, the specific star formation rates, SSFR = SFR/$M_{\rm *}$,
and star formation rate densities, $\dot{\rho}_{\rm *}$, 
we infer from the hard and soft band emission of the late-type galaxies
span the range of previously reported values (Figs.~\ref{hspecSSFR} and~\ref{hspecSFRd}).

The specific X-ray luminosities we found for the early-type galaxies are in good agreement with 
the results of past studies (O'Sullivan et al. 2001, G03, B05, and Lehmer 2007).
Additionally, our work provides the first X-ray estimates of the SMBH accretion rates, 
$\dot{M}_{\rm BH}$, and accretion rate densities, $\dot{\rho}_{\rm BH}$, in normal galaxies at these redshifts.
Our findings suggest that the redshift evolution of
low luminosity AGN in our early-type galaxy sample
is similar to that of
higher luminosity AGN (Ba01 and Ba05), which we exclude from our analysis.  However, the lower
luminosity AGN contribute little to the overall growth of supermassive black holes.
In general, our work shows that there is a continuum rather than a sudden break
in the star formation and SMBH accretion histories of galaxies from the 
powerful starbursts and AGN of the past to the fainter,
optically-normal galaxies more prevalent today.

Our analysis of the X-ray evolution of galaxies in the NDWFS XBo\"otes field can be significantly
expanded in the future. A second phase of AGES has doubled the size of the spectroscopic sample
and extended the redshift range of galaxies.  If combined with photometric redshifts
based on the very extensive, multiwavelength NDWFS photometry, it will be possible to
reach $z \sim 1$ to a uniform luminosity limit and greatly reduce statistical errors.

\begin{acknowledgments}
CRW thanks John Beacom, Mark Brodwin, Xinyu Dai, Stephan Frank, Oleg Gnedin, Andy Gould, Dirk Grupe, Himel Ghosh, 
Matt Kistler, Smita Mathur, Pat Osmer, Rick Pogge, Ken Rines, Louie Strigari, Ezequiel Treister, Meg Urry, 
Pieter van Dokkum, Jeff van Duyne, David Weinberg, Jong-Hak Woo and Hasan Yuksel for helpful discussions. We also
thank the referee for useful and extensive comments.

This research was supported by the National Optical Astronomy 
Observatory which is operated by the Association of 
Universities for Research in Astronomy (AURA), Inc.$_{~}$under
a cooperative agreement with the National Science 
Foundation. Spectroscopic observations reported here were 
obtained at the MMT Observatory, a joint facility of the 
Smithsonian Institution and the University of Arizona.
We thank the CXC for scheduling the CXO observations
that made this work possible and the CXC Data Processing Team for the pipeline data.
\end{acknowledgments}

\end{document}